\documentclass[aps,prb,twocolumn, a4paper,amsmath,amssymb,superscriptaddress]{revtex4-2}

\usepackage[colorlinks,citecolor=blue,urlcolor=blue]{hyperref}
\usepackage{txfonts}
\usepackage{graphicx}
\usepackage{dcolumn}
\usepackage{bm}
\usepackage[usenames]{color} 
\usepackage{amssymb} 
\usepackage{amsmath} 
\usepackage[utf8]{inputenc} 
\usepackage{braket}
\usepackage{mathptmx}
\usepackage{physics}
\usepackage{siunitx}

\newcommand{\Uell}{{u}_{a,{\bm k}}}
\newcommand{\Uellb}{{u}_{b {\bm k}}}

\newcommand{\Eell}{\varepsilon_{a}}
\newcommand{\Eellb}{\varepsilon_{b}}

\newcommand{\Eellx}{\partial_x \Eell}
\newcommand{\Eelly}{\partial_y \Eell}
\newcommand{\Eellm}{\partial_\mu \Eell}
\newcommand{\Eelln}{\partial_\nu \Eell}

\begin{document}

\title{Effects of Berry Curvature and Orbital Magnetic Moment in the Magnetothermoelectric Transport of Bloch Electron Systems}

\date{\today}

\author{Viktor \surname{K\"{o}nye}}
\affiliation{Institute for Theoretical Physics Amsterdam, University of Amsterdam, Science Park904, 1098 XH Amsterdam, The Netherlands}
\affiliation{Institute for Theoretical Solid State Physics, IFW Dresden and W\"urzburg-Dresden Cluster of Excellence ct.qmat, Helmholtzstr. 20, 01069 Dresden, Germany}
\author{Masao \surname{Ogata}}
\affiliation{
 Department of Physics, University of Tokyo, Hongo, Bunkyo-ku, Tokyo 113-0033, Japan
\\
and Trans-scale Quantum Science Institute, University of Tokyo, Hongo, Bunkyo-ku, Tokyo 113-0033, Japan
}

\begin{abstract}
Thermoelectric transport coefficients up to linear order in the applied magnetic field are microscopically studied using Kubo-Luttinger linear response theory and thermal Green's functions. 
We derive exact formulas for the thermoelectric conductivity and thermal conductivity in the limit of small relaxation rates for Bloch electrons in terms of Bloch wave functions, which show that the Sommerfeld-Bethe relationship holds.
Our final formula contains the Berry curvature contributions as well as the orbital magnetic moment contributions, that arise naturally from the microscopic theory.
We show that generalized $f$-sum rules containing the Berry curvature and orbital magnetic moment play essential roles in taking into account the interband effects of the magnetic field.
As an application, we study a model of a gapped Dirac electron system with broken time-reversal symmetry and show the presence of a linear magnetothermopower in such systems. 
\end{abstract}

\maketitle

\section{Introduction}

Inspired by the anomalous velocity \cite{Niu} related to the Berry curvature in the semiclassical Boltzmann transport equation \cite{Ziman,Solyom}, the contributions of the Berry curvature in transport properties \cite{KarplusLuttinger,TKNN,Kohmoto,Cortijo2016,Gao2017,Zyuzin2017,Nandy2018,Sun2019,Ma2019,Das2019,Konye2020}, orbital magnetic moment \cite{Xiao2005,Thonhauser,Thonhauser2,ShiVignale}, orbital magnetic susceptibility \cite{Gao,Raoux,Ogata2017}, and orbital-Zeeman susceptibility \cite{Nomura,Ozaki,Ozaki2} have been extensively studied. 

In addition to the quantum Hall effect \cite{TKNN,Kohmoto} the Berry curvature contribution to the anomalous Hall effect was already obtained by Karplus and Luttinger microscopically using perturbation theory \cite{KarplusLuttinger},
although at that time it was not called {\it Berry curvature}.
Similarly, in the case of the orbital magnetic susceptibility, the Berry curvature terms were included in the studies by Hebborn and Sondheimer \cite{Hebborn} and by Blount \cite{Blount}.
Actually, the notation of $\bm \Omega$ for the Berry curvature originates from the paper by Blount \cite{Blount}.

The Berry curvature vanishes in the presence of both inversion symmetry and time-reversal symmetry.
In the case of time-reversal symmetry breaking, the typical example of the Berry curvature contribution to the conductivity is the Hall conductivity $\sigma_{xy}$ in the quantum Hall effect \cite{TKNN,Kohmoto} and anomalous Hall effect \cite{KarplusLuttinger}.
In contrast, in the diagonal components of the conductivity tensor such as $\sigma_{xx}$ or $\sigma_{zz}$, anomalous contributions proportional to the applied magnetic field $B$ can appear due to the Berry curvature \cite{Cortijo2016,Gao2017,Zyuzin2017,Nandy2018,Sun2019,Ma2019,Das2019,Konye2020}.
[Note that conventional magnetoconductivity is proportional to $B^2$.]

Furthermore, it has been discussed that the Seebeck coefficient and figure of merit $ZT=S^2 \sigma T/\kappa$ can also be enhanced in the presence of a magnetic field \cite{wolfe1962,skinner2018,konye2019}. Therefore, a microscopic understanding of thermoelectric conductivity in the presence of a magnetic field is also important for applications such as thermoelectric devices.

Recently, we calculated the magnetoconductivity in terms of thermal Green’s functions \cite{Konye2020} based on Fukuyama's formula \cite{Fukuyama1969} and found that there are contributions from the Berry curvature $\Omega^a_{\mu\nu}$ and orbital magnetic moment $M^a_{\mu\nu}$ in $\sigma_{\mu\nu}(B)$ in the linear order of $B$.
Here, $a$ represents the band index and $\mu, \nu = x, y$, or $z$.
The orbital magnetic moment $M^a_{\mu\nu}$ is expressed by [see Eq.~(\ref{eq:MmunuDef}) below] which is absent in the Boltzmann transport theory with anomalous velocities.
To discuss the transport properties, microscopic calculations using the Kubo’s linear response theory and thermal Green’s functions are very important.
Similar situation also appeared in the studies of orbital magnetic susceptibility, $\chi_{\rm orb}$. 
Gao et al. \cite{Gao} studied $\chi_{\rm orb}$ using the semiclassical Boltzmann transport theory with Berry curvature and found that $\chi_{\rm orb}$ can be expressed in terms of various contributions including the Berry curvature contributions. 
However, it has been shown that some of the obtained coefficients of the contributions do not agree with the microscopically obtained results \cite{Hebborn,Ogata2017}.

In this paper, we extend our previous theory for the conductivity \cite{Konye2020} to thermoelectric and thermal conductivities.
When we use the classical Boltzmann transport theory, the following relation between the transport coefficients $L_{11}$, $L_{12}$, and $L_{22}$ holds 
\begin{equation}\begin{split}
L_{11}&= \int_{-\infty}^\infty d\varepsilon \left(-f'(\varepsilon)\right) \sigma(\varepsilon, T), \cr
L_{12}=L_{21}&= \frac{1}{e} \int_{-\infty}^\infty d\varepsilon (\varepsilon-\mu) \left(-f'(\varepsilon) \right) \sigma(\varepsilon, T), \cr
L_{22}&= \frac{1}{e^2} \int_{-\infty}^\infty d\varepsilon (\varepsilon-\mu)^2 
\left(-f'(\varepsilon)\right) \sigma(\varepsilon, T), 
\label{eq:SBRel}
\end{split}\end{equation}
which we call Sommerfeld-Bethe relation. Here $e(<0)$ is the charge of electron and
$f(\varepsilon)$ is the Fermi distribution function,
$f(\varepsilon)=1/[e^{\beta (\varepsilon-\mu)}+1]$, with $\beta=1/k_{\rm B}T$. 
One of the issues to be studied is that whether or not this kind of relation holds under the magnetic field and
for the contributions from the Berry curvature that does not have $f'(\varepsilon)$ but $f(\varepsilon)$. 
In this paper, we derive $L^{12}_{\mu\nu}$ and $L^{22}_{\mu\nu}$ up to linear order of the external magnetic field using Kubo-Luttinger linear response theory and thermal Green’s functions. 
Compared with the calculation of the magnetic susceptibility \cite{Ogata2015,Ogata2017} that is a thermodynamic quantity, in the present calculation of $L^{ij}_{\mu\nu}$, the finite frequency Green's functions are to be carefully taken into account since $L^{ij}_{\mu\nu}$ is a dynamical quantity. 

In section II, we introduce our model and define the current and heat current 
operators, $\bm j(\bm r)$ and $\bm j_Q (\bm r)$. 
Using the imaginary time interaction representation, we show a specific 
relationship between $\bm j(\bm r, \tau, \tau')$ and $\bm j_Q (\bm r, \tau, \tau')$, 
which will be used later.
In section III, we define the electronic and thermoelectric linear response functions 
$L^{ij}_{\mu\nu}$ represented by the analytic continuation of the correlation functions. 
We show our final results in Eq.~(\ref{eq:LmunuFinal}) before describing the details of the calculation. 
In section IV, we calculate the correlation functions in the linear order of the magnetic field.
Since we use the general Bloch band representation, there are seven contributions depending 
on the interband matrix elements of the current operator as shown in Fig.~1. 
Then, we calculate each contribution by taking the leading and subleading order 
with respect to $\Gamma^\ell$ ($\ell=-2$ and $-1$) assuming that the relaxation rate $\Gamma$
of electrons is small. 
The results in the leading order of $O(\Gamma^{-2})$ are shown in section IV A, where 
we reproduce the results in the Boltzmann theory for the Hall conductivity. 
We also derive a generalized $f$-sum rule that contains the orbital magnetic moment, which 
is used for summing up the interband contributions. 
The results in the subleading order of $O(\Gamma^{-1})$ are summarized in section IV B, 
where another generalized $f$-sum rule containing the Berry curvature is derived and used.
In section V, we study a simple two-band model that shows an anomalous linear-$B$ behavior 
of the diagonal components of the linear response functions. 
We show the Seebeck coefficient, thermal conductivity, Lorenz number, power factor, and 
figure of merit $zT$ for this model. 
We summarize our results in Section VI.
The details of the calculations are shown in Appendices. 

\section{Model, current, and heat current operators}

We consider Bloch electrons in a periodic potential $V({\bm r})$ 
and spin-orbit interactions using the following Hamiltonian 
derived from the Dirac equation with non-relativistic expansion:
\begin{equation}
\begin{split}
{\mathcal H} 
&=\int d{\bm r} \psi^\dagger({\bm r}) H ({\bm r})\psi({\bm r}), \cr
        H({\bm r}) &=
\frac{1}{2m}{\bm p}^2 
+ V({\bm r}) +\frac{\hbar^2}{8m^2c^2} {\bm \nabla}^2 V({\bm r}) 
- \frac{e\hbar}{2m} {\bm \sigma}\cdot {\bm b}({\bm r})\cr
&+ \frac{\hbar}{4m^2c^2} {\bm \sigma}\cdot {\bm \nabla} V \times {\bm p},
\label{Hamiltonian}
\end{split}
\end{equation}
where ${\bm p} = -i\hbar {\bm \nabla}$, 
$\sigma_i$ are the Pauli matrices representing spin, 
and $\psi({\bm r})$ ($\psi^\dagger({\bm r})$) is the two-component 
electron annihilation (creation) field operator.  
An effective field ${\bm b}({\bm r})$ is phenomenologically introduced which 
breaks time-reversal symmetry. 
The last term represents the spin-orbit interaction in a general form. 
We treat the general case where $V({\bm r})$ is a noncentrosymmetric 
potential (i.e., $V(-{\bm r})\ne V({\bm r})$), similarly to Ref.~\cite{Ogata2017}.

We take the Bloch wave functions in the form $e^{i{\bm k}\cdot {\bm r}} \Uell({\bm r})$, which satisfy
\begin{equation}
H_{\bm k} \Uell({\bm r}) = \Eell({\bm k}) \Uell({\bm r}).
\label{UellEq}
\end{equation}
Here, $a$ indicates the band index including the pseudospin degrees of freedom, and 
$H_{\bm k}:= e^{-i\bm k\cdot \bm r}  H(\bm r) e^{i\bm k\cdot \bm r}$
is given by
\begin{equation}\begin{split}
H_{\bm k} &= \frac{\hbar^2 k^2}{2m} - \frac{i\hbar^2}{m} {\bm k}\cdot {\bm \nabla}
- \frac{\hbar^2}{2m} {\bm \nabla}^2 + V({\bm r})
+\frac{\hbar^2}{8m^2c^2} {\bm \nabla}^2 V \cr
&-\frac{e\hbar}{2m} {\bm \sigma}\cdot {\bm b} 
+ \frac{\hbar^2}{4m^2c^2} {\bm \sigma}\cdot {\bm \nabla} V \times 
\left( {\bm k} - i{\bm \nabla}\right),
\label{HamiltonianK}
\end{split}\end{equation}
in the present Hamiltonian of Eq.~(\ref{Hamiltonian}).
Note that the function $\Uell({\bm r})$ is a two-component vector (representing the 
spin degrees of freedom), which is periodic with the same period as $V({\bm r})$.

The current operator $\bm j(\bm r)$ and the energy current operator $\bm j_E(\bm r)$
are defined via the continuity equations \cite{Mahan1990}
\begin{equation}\begin{split}
\frac{d}{dt} e\rho(\bm r) + {\rm div} \bm j(\bm r) &=0, \cr
\frac{d}{dt} h(\bm r) + {\rm div} \bm j_E(\bm r) &=0,
\label{eq:continuity}
\end{split}\end{equation}
where $\rho(\bm r)=\psi^\dagger(\bm r) \psi(\bm r)$ represents the electron density
operator  
and $h(\bm r)=\frac{1}{2} \psi^\dagger(\bm r) \{  H(\bm r) 
+{\overleftarrow H}(\bm r)\} \psi(\bm r)$ is the symmetrized Hamiltonian density
operator. 
Here, $\overleftarrow H(\bm r)$ is defined by 
replacing the operator $p_j = -i\hbar \partial_j$ ($j=x, y, z$) in 
$ H(\bm r)$ with 
$\overleftarrow {p_j} := i\hbar \overleftarrow {\partial_j}$, where 
$\overleftarrow {\partial_j}$ represents the partial derivative with respect to $j$ 
that acts on the functions and operators on its left-hand side. 
The heat current operator is given by 
$\bm j_Q (\bm r) = \bm j_E (\bm r) -\mu \bm j(\bm r)/e$
with $\mu$ being the chemical potential.

Using the continuity equations, we obtain
\begin{equation}\begin{split}
\bm j(\bm r)&=e\psi^\dagger(\bm r)\overleftrightarrow{\bm j_0}(\bm r)\psi(\bm r), \cr
\bm j_{E} (\bm r)&= \frac{1}{2} \psi^\dagger(\bm r) \left\{ 
\overleftrightarrow{\bm j_0}(\bm r)  H (\bm r)
+ \overleftarrow H (\bm r) \overleftrightarrow{\bm j_0}(\bm r)\right\} \psi(\bm r),
\label{eq:Jtot}
\end{split}\end{equation}
with
\begin{equation}
\overleftrightarrow {{\bm j}_0}({\bm r})
=\frac{1}{2m} (-i\hbar {\bm \nabla} + i\hbar \overleftarrow{\bm \nabla})
+ \frac{\hbar}{4m^2c^2} {\bm \sigma}\times {\bm \nabla}V.
\label{eq:jdefs}
\end{equation}
The derivation including the case with the vector potential is given in Appendix A.
We can show that the heat current and the electric current operator are related to each other \cite{JonsonMahan,OgataFukuyama2019}.
Defining the imaginary time interaction representation of the electric current operator as
\begin{equation}
{\bm j}({\bm r},\tau,\tau') = e\psi^\dagger({\bm r},\tau) 
\overleftrightarrow {\bm j_0} ({\bm r}) \psi({\bm r}, \tau'),
\end{equation}
with $\psi^\dagger({\bm r},\tau) =e^{\tau({\mathcal H}-\mu N)}  \psi^\dagger({\bm r})
e^{-\tau ({\mathcal H}-\mu N)}$ and 
$\psi({\bm r},\tau') =e^{\tau' ({\mathcal H}-\mu N)}  \psi({\bm r})
e^{-\tau' ({\mathcal H}-\mu N)}$, 
it can be shown that
\begin{equation}
\frac{1}{2e} \left( \frac{\partial}{\partial \tau} - \frac{\partial}{\partial \tau'} \right)
{\bm j} ({\bm r},\tau,\tau')
= {\bm j}_{Q}({\bm r},\tau,\tau'),
\label{JMrelation}
\end{equation}
where ${\bm j}_{Q}({\bm r},\tau,\tau' )$ is the similarly-defined imaginary time 
interaction representation of the heat current operator, 
${\bm j}_{Q}({\bm r})=\bm j_{E}(\bm r)-\mu \bm j(\bm r)/e$. (see also Appendix A.)
This type of relationship was used by Jonson and Mahan \cite{JonsonMahan} to show 
the Sommerfeld-Bethe relationship (\ref{eq:SBRel}) between the electric conductivity and thermoelectric 
conductivity \cite{OgataFukuyama2019}. Note that, when there is electron-phonon interaction or 
finite-range mutual interaction between electrons, the relation in 
Eq.~(\ref{JMrelation}) does not hold \cite{OgataFukuyama2019}.
In the present Hamiltonian (\ref{Hamiltonian}), 
Eq.~(\ref{JMrelation}) does hold even in the presence of the spin-orbit interaction.
In the following sections, we will show that the Sommerfeld-Bethe relationship of Eq.~(\ref{eq:SBRel}) holds even in the presence of a magnetic field, where the heat current operator also depends on the magnetic field.

The current density in terms of the Bloch wave functions is
\begin{equation}
(j_\mu)_{ab}= e\int u_{a, \bm k}^\dagger (\bm r) e^{-i\bm k\cdot \bm r} 
\overleftrightarrow{j_{0,\mu}} (\bm r) e^{i\bm k\cdot \bm r} u_{b,\bm k}(\bm r) d\bm r.
\end{equation}
We define $(\gamma_\mu)_{ab}=\frac{\hbar}{e} (j_\mu)_{ab}$.
Using the definition of $H_{\bm k}$ in Eq.~(\ref{HamiltonianK}) and the $k_\mu$-derivative of $(\varepsilon_a - H) |a\rangle =0$, we obtain
\begin{equation}\begin{split}
(\gamma_\mu)_{ab} &= \langle a|\partial_\mu H | b\rangle = \partial_\mu \varepsilon_{a} \delta_{ab} + p_{ab,\mu}, \cr
p_{ab,\mu} &:= (\varepsilon_b - \varepsilon_a )\langle a | \partial_\mu b \rangle,
\label{eq:gammaDef}
\end{split}\end{equation}
where $|a \rangle = u_{a,{\bm k}}({\bm r})$, 
$\partial_\mu H = \partial H_{\bm k}/\partial k_\mu$, 
$\partial_\mu \varepsilon_a = \partial \varepsilon_a/\partial k_\mu$, 
$|\partial_\mu a \rangle = \partial u_{a,{\bm k}}({\bm r}) / \partial k_\mu$,
and the wave number $\bm k$ dependence is not shown explicitly, 
since it is common in all the expressions.
Apparently, $p_{ab,\mu}$ is closely related to the interband Berry connection.

\section{Transport coefficients}

The electric and thermoelectric linear response functions are defined by
\begin{equation}\begin{split}
\langle j_\mu \rangle &= \sum_\nu
L^{11}_{\mu\nu} E_\nu + L^{12}_{\mu\nu} \left( -\frac{\partial_\nu T}{T} \right), \cr
\langle j_{Q,\mu} \rangle &= \sum_\nu
L^{21}_{\mu\nu} E_\nu + L^{22}_{\mu\nu} \left( -\frac{\partial_\nu T}{T} \right),
\end{split}\end{equation}
for $\mu, \nu = x,y,z$, where
$L^{11}_{\mu\nu}$ is the usual electric conductivity tensor, $L^{12}_{\mu\nu}$ and 
$L^{21}_{\mu\nu}$ are the thermoelectric and electrothermal conductivity, 
and the thermal conductivity is obtained from
$L^{22}_{\mu\nu}$. 
The observable quantities can be computed as
\begin{equation}
\label{eq:transport}
\begin{split}
        \sigma &= \vb{L}_{11},\\
        S &= \frac{1}{T}\vb{L}_{11}^{-1}\vb{L}_{12},\\
        \kappa &= \frac{1}{T}\left[\vb{L}_{22}-\vb{L}_{21}\vb{L}_{11}^{-1}\vb{L}_{12}\right]
\end{split}
\end{equation}
where $\sigma$ is the electric conductivity tensor, $S$ is the Seebeck tensor and $\kappa$ is the thermal conductivity tensor.

In the linear response theory or in the Kubo formula, the conductivity $L^{11}_{\mu\nu}$ is obtained as
\begin{equation}
L^{11}_{\mu\nu} = \lim_{\omega\rightarrow 0} \frac{1}{i(\omega+i\delta)} 
\left\{ \tilde\Phi^{11}_{\mu\nu}(\omega+i\delta) - \tilde\Phi^{11}_{\mu\nu} (0) \right\}, 
\label{eq:Conduc}
\end{equation}
where, $\tilde\Phi^{11}_{\mu\nu} (\omega+i\delta)$ is the analytic continuation of the current-current correlation function 
\begin{equation}
\Phi^{11}_{\mu\nu} (i\omega_\lambda) = \frac{1}{V} \int_0^\beta d\tau 
\langle j_{\bm k={\bm 0},\mu} (\tau) j_{\bm k={\bm 0},\nu} (0)\rangle 
e^{i\omega_\lambda \tau}, 
\label{ThermalCorr}
\end{equation}
with $i\omega_\lambda \rightarrow \hbar(\omega + i\delta)$ and 
$\delta$ being an infinitesimal small parameter. 
Here $\beta=1/k_{\rm B}T$, $\omega_\lambda = 2\pi \lambda k_{\rm B}T$ is the Matsubara
frequency with $\lambda\in\mathbb{Z}$, and 
$j_{{\bm k},\mu}(\tau)$ is the $\mu$-component of the Fourier transform of 
the current density operator defined by
\begin{equation}
{\bm j}_{\bm k} (\tau) = \int d{\bm r} {\bm j} ({\bm r}, \tau,\tau) e^{-i{\bm k}\cdot {\bm r}}.
\end{equation}

The other linear response coefficients, $L^{12}_{\mu\nu}=L^{21}_{\mu\nu}$ and $L^{22}_{\mu\nu}$ are similarly obtained using the heat current operators. 
We calculate them in the linear order of the applied magnetic field $B$ in the $z$-direction. 
Since the calculation is lengthy and complicated, we show our final results beforehand. 
Expressing $L_{\mu\nu}^{11}, L_{\mu\nu}^{12},$ and $L_{\mu\nu}^{22}$ as $L_{\mu\nu}^{(0)}, L_{\mu\nu}^{(1)},$ and $L_{\mu\nu}^{(2)}$, respectively, we obtain
\begin{equation}\begin{split}
L_{\mu\nu}^{(\ell)} &= -\frac{e^{3-\ell} B}{\hbar^2 V}\sum_{{\bm k},a} 
(\varepsilon_a-\mu)^\ell f'(\varepsilon_a)\cr
\times &\biggl[ \frac{1}{8\Gamma^2}
\bigl\{ \Eellm \Eellx (\partial_\nu \partial_y \Eell) 
 - \Eellm \Eelly (\partial_\nu \partial_x \Eell) \cr
&- \Eelln \Eellx (\partial_\mu \partial_y \Eell) 
 + \Eelln \Eelly (\partial_\mu \partial_x \Eell) \bigr\} \cr
&-\frac{1}{2\Gamma} ( \Eellm \Eelln \Omega_{xy}^a 
 - \Eellm \Eellx \Omega_{\nu y}^a + \Eellm \Eelly \Omega_{\nu x}^a \cr
&- \Eelln \Eellx \Omega_{\mu y}^a + \Eelln \Eelly \Omega_{\mu x}^a )\cr
&-\frac{1}{4\Gamma} \left( \partial_\mu \Eell \partial_\nu M_{xy} 
+ \partial_\nu \Eell \partial_\mu M_{xy} -2(\partial_\mu \partial_\nu \Eell) 
M_{xy} \right) \biggr] \cr
&+O(\Gamma^0), \quad (\ell = 0,1,2),
\label{eq:LmunuFinal}
\end{split}\end{equation}
where $\Gamma$ is the relaxation rate of electrons and $\partial_\mu$ represents the partial derivative $\partial/\partial k_\mu$. 
Here, we calculate the transport coefficients $L_{\mu\nu}^{(\ell)}$ in the power expansion with respect to $\Gamma^{-n}$ assuming $\Gamma$ is small compared with the Fermi energy. 
The contribution proportional to $1/\Gamma^2$ is exactly the same as the result of the Boltzmann transport theory. 
The contributions proportional to $1/\Gamma$ are related to the Berry curvature for the band $a$
\begin{equation}
\Omega_{\mu\nu}^a 
:= -2{\rm Im} 
\langle \partial_\mu a |\partial_\nu a \rangle
= i \left\{ \langle \partial_\mu a |\partial_\nu a \rangle - \langle \partial_\nu a |\partial_\mu a \rangle \right\}.
\label{eq:OmegamunuDef}
\end{equation}
as well as the orbital magnetic moment
\begin{equation}\begin{split}
M_{\mu\nu}^a :=& {\rm Im} \langle \partial_\mu a | \Eell- H_{\bm k} | \partial_\nu a \rangle \cr
=&\frac{1}{2i} \left( 
\langle \partial_\mu a | \Eell- H_{\bm k} | \partial_\nu a \rangle 
-\langle \partial_\nu a | \Eell- H_{\bm k} | \partial_\mu a \rangle \right).
\label{eq:MmunuDef}
\end{split}\end{equation}
When we define 
\begin{equation}\begin{split}
\sigma(\varepsilon, T) &= -\frac{e^{3} B}{\hbar^2 V}\sum_{{\bm k},a} \delta(\varepsilon-\varepsilon_a) \cr
\times &\biggl[ \frac{1}{8\Gamma^2}
\bigl\{ \Eellm \Eellx (\partial_\nu \partial_y \Eell) 
 - \Eellm \Eelly (\partial_\nu \partial_x \Eell) \cr
&- \Eelln \Eellx (\partial_\mu \partial_y \Eell) 
 + \Eelln \Eelly (\partial_\mu \partial_x \Eell) \bigr\} \cr
&-\frac{1}{2\Gamma} ( \Eellm \Eelln \Omega_{xy}^a 
 - \Eellm \Eellx \Omega_{\nu y}^a + \Eellm \Eelly \Omega_{\nu x}^a \cr
&- \Eelln \Eellx \Omega_{\mu y}^a + \Eelln \Eelly \Omega_{\mu x}^a )\cr
&-\frac{1}{4\Gamma} \left( \partial_\mu \Eell \partial_\nu M_{xy} 
+ \partial_\nu \Eell \partial_\mu M_{xy} -2(\partial_\mu \partial_\nu \Eell) 
M_{xy} \right) \biggr],
\end{split}\end{equation}
we can easily see that the Sommerfeld-Bethe relation in Eq.~(\ref{eq:SBRel}) holds.

Let us remark here a problem in transverse thermoelectric transport \cite{Smrcka1977,cooper1997,qin2011,shitade2014,dyrdal2016,fujimoto2024}.
In the case of transverse thermoelectric transport, a term appears which diverges as $1/T$ as $T\rightarrow 0$.
This nonphysical divergence is removed by considering the local equilibrium current \cite{fujimoto2024}, which is in the order of $\Gamma^0$ since the local equilibrium current does not depend on the relaxation.
Although the same problem will appear in the present formalism, it only appears at higher orders than the ones discussed in this paper $\Gamma^{-2}$ and $\Gamma^{-1}$, as shown in Eq.~(\ref{eq:LmunuFinal}).

\section{Correlation functions in the linear order of the magnetic field}

In the presence of a magnetic field, Fukuyama \cite{Fukuyama1969A,Fukuyama1969} obtained an exact formula for the Hall conductivity or $\Phi_{xy}^{11}(i\omega_\lambda)$
expressed in terms of thermal Green's functions in a 
gauge-invariant form using a finite momentum vector potential 
$\bm A(\bm r) = e^{i\bm q\cdot \bm r} \bm A_{\bm q}$.
Recently in Ref.~\cite{Konye2020} we simplified the formula of $\Phi_{xy}^{11}(i\omega_\lambda)$ and furthermore calculated the formula for the other components such as $\Phi_{zz}^{11}(i\omega_\lambda)$. 
In the present paper, we study $\Phi_{\mu\nu}^{11}(i\omega_\lambda)$, $\Phi_{\mu\nu}^{12}(i\omega_\lambda)$, and $\Phi_{\mu\nu}^{22}(i\omega_\lambda)$ for all the components $\mu, \nu = x,y,z$ in a unified way. 
As shown in Appendix B, the correlation function $\Phi_{\mu\nu}^{11}(i\omega_\lambda)$ in the linear order of the magnetic field can be simplified as
\begin{equation}\begin{split}
\Phi_{\mu\nu}^{11} (i\omega_\lambda) &= -\frac{k_{\rm B}T}{V} \sum_{n,{\bm k}} 
\frac{ie^3 B}{\hbar^3} {\rm Tr}\ F(\bm k, i\varepsilon_n, i\omega_\lambda) \cr
F(\bm k, i\varepsilon_n, i\omega_\lambda)
&= \frac{\hbar^2}{2m} \bigl[ 
\delta_{\mu y} ({\mathcal G}_+ \gamma_\nu {\mathcal G}\gamma_x {\mathcal G} 
-{\mathcal G}_+ \gamma_x {\mathcal G}_+ \gamma_\nu {\mathcal G}) \cr
&+\delta_{\nu x} (\gamma_\mu {\mathcal G}_+ {\mathcal G} \gamma_y {\mathcal G}
-\gamma_\mu {\mathcal G}_+ \gamma_y {\mathcal G}_+ {\mathcal G}) \bigr] \cr
&+ \gamma_\mu {\mathcal G}_+ \gamma_\nu {\mathcal G} \gamma_x {\mathcal G} \gamma_y {\mathcal G} 
-\gamma_\mu {\mathcal G}_+ \gamma_y {\mathcal G}_+ \gamma_x {\mathcal G}_+ \gamma_\nu {\mathcal G},
\label{Ogata2019}
\end{split}\end{equation}
where the magnetic field $B$ is applied in the $z$-direction without loss of generality.  
The summation of the wave vector $\bm k$ is taken within the first Brillouin zone, 
the summation of $n$ represents the Matsubara frequency summation, 
and the trace (Tr) is taken over all the eigenstates of the Bloch states including the 
spin degrees of freedom.
Note that the simplified formula for $\Phi_{xy}^{11}(i\omega_\lambda)$ enables us to show the close relationship between the orbital magnetic susceptibility and a part of the Hall conductivity \cite{ogata2024}.

The thermal Green's function $\mathcal G$ is band-index diagonal, which has a form 
\begin{equation}
{\mathcal G}_a({\bm k},\varepsilon_n)=\frac{1}{i\varepsilon_n-\Eell({\bm k})+\mu 
+i\Gamma {\rm sign} (\varepsilon_n)},
\end{equation}
with $\varepsilon_n=(2n+1)\pi k_{\rm B}T$ being the fermion Matsubara frequency, 
and we have abbreviated as 
${\mathcal G}({\bm k}, i\varepsilon_n) \rightarrow {\mathcal G}$, 
${\mathcal G}({\bm k}, i\varepsilon_n+i\omega_\lambda) \rightarrow {\mathcal G}_+$.
Here, we have assumed the case with simple random impurities that have 
$\delta$-function potentials, for simplicity. 
In this case, the self-energy can be approximated as 
$-i\Gamma{\rm sign}(\varepsilon_n)$ and the vertex 
corrections due to impurity scattering are safely neglected. 
This is the simplest assumption and, for actual models, we have to modify the 
self-energy depending on the explicit model, which remains as an extension of the 
present theory. 
In the following, we evaluate the transport coefficients in the expansion with 
respect to $\Gamma^n$ assuming that $\Gamma<<\varepsilon_{\rm F}$.

In Ref.~\cite{Konye2020} we calculated the conductivity tensors $L^{11}_{\mu\nu}$ using 
Eq.~(\ref{Ogata2019}). 
In the present paper, we discuss $L^{12}_{\mu\nu}$, $L^{21}_{\mu\nu}$, and $L^{22}_{\mu\nu}$. 
For example, the electrothermal conductivity $L^{21}_{\mu\nu}$ can be obtained from 
the analytic continuation ($i\omega_\lambda \rightarrow \hbar(\omega + i\delta)$) of
\begin{equation}
\Phi^{21}_{\mu\nu} (i\omega_\lambda) = \frac{1}{V} \int_0^\beta d\tau 
\langle j_{Q,\bm k={\bm 0},\mu} (\tau) j_{\bm k={\bm 0},\nu} (0)\rangle 
e^{i\omega_\lambda \tau},
\label{ThermalCorr2}
\end{equation}
in which $j_{Q,\bm k,\mu} (\tau)$ is defined as 
\begin{equation}
{\bm j}_{Q, \bm k} (\tau) = \int d{\bm r} {\bm j}_Q ({\bm r}, \tau,\tau) e^{-i{\bm k}\cdot {\bm r}},
\end{equation}
Because ${\bm j}_Q ({\bm r}, \tau,\tau)$ is related to the current operator 
${\bm j} ({\bm r}, \tau,\tau)$ as in Eq.~(\ref{JMrelation}),
we can show that each Feynman diagram to calculate Eq.~(\ref{ThermalCorr2})
contains the $\tau$-derivatives of the thermal Green's functions. 
(Note that the relation in Eq.~(\ref{JMrelation}) also holds even in the presence of 
magnetic field, as shown in Appendix A.)
Since the $\tau$-derivatives yield the Matsubara frequencies of the 
Green's functions, we find
\begin{equation}\begin{split}
\Phi_{\mu\nu}^{12} (i\omega_\lambda)&= -\frac{k_{\rm B}T}{V} \sum_{n,{\bm k}} 
\frac{ie^2 B}{\hbar^3} 
\biggl( i\varepsilon_n + \frac{i\omega_\lambda}{2} \biggr) 
{\rm Tr}\ F(\bm k, i\varepsilon_n, i\omega_\lambda), \cr
\Phi^{21}_{\mu\nu} (i\omega_\lambda) &=\Phi^{12}_{\mu\nu}(i\omega_\lambda), \cr
\Phi_{\mu\nu}^{22} (i\omega_\lambda)&= -\frac{k_{\rm B}T}{V} \sum_{n,{\bm k}} 
\frac{ie B}{\hbar^3} 
\biggl( i\varepsilon_n + \frac{i\omega_\lambda}{2} \biggr)^2
{\rm Tr} \ F(\bm k, i\varepsilon_n, i\omega_\lambda).
\label{PhiJJQ0}
\end{split}\end{equation}
Note that the additional factor $(i\varepsilon_n+i\omega_\lambda/2)^\ell$ 
($\ell=1,2$) does not depend on the wave numbers, so that the 
gauge-invariance argument in Ref. \cite{Fukuyama1969} that uses the $\bm k$-dependence of Green's functions and current operators are not affected.

Similarly to the conductivity expressed in Ref.~\cite{Konye2020} we express the transport coefficients as
\begin{equation}\begin{split}
L^{(\ell)}_{\mu\nu} &= \frac{e^{3-\ell} B}{\hbar^2 V} \sum_{{\bm k}}  \cr
&\phantom{+}\frac{\hbar^2}{2m}\delta_{\mu y} \left\{ (\gamma_\nu)_{ab} (\gamma_x)_{ba} C^{(\ell)}_{aba} 
-(\gamma_x)_{ab} (\gamma_\nu)_{ba} \tilde C^{(\ell)}_{aba} \right\} \cr
&+\frac{\hbar^2}{2m}\delta_{\nu x} \left\{(\gamma_y)_{ab} (\gamma_\mu)_{ba} C^{(\ell)}_{aab}
-(\gamma_\mu)_{ab} (\gamma_y)_{ba} \tilde C^{(\ell)}_{baa}\right \} \cr
&+ (\gamma_\mu)_{da} (\gamma_\nu)_{ab} (\gamma_x)_{bc} (\gamma_y)_{cd} D^{(\ell)}_{abcd}\cr 
&-(\gamma_\mu)_{ad} (\gamma_y)_{dc} (\gamma_x)_{cb} (\gamma_\nu)_{ba} \tilde D^{(\ell)}_{dcba},
\label{eq:L11withD}
\end{split}\end{equation}
where $\ell$ is defined using the indices of the transport coefficients $\vb{L}_{ij}$ as $\ell=i+j-2$.
The $C^{(\ell)}_{abc}$ $D^{(\ell)}_{abcd}$  ($\ell=0,1,2$) functions are defined as
\begin{equation}\begin{split}
&C^{(\ell)}_{abc} 
= -\lim_{\omega\rightarrow 0} \frac{k_{\rm B}T}{\hbar\omega} \sum_{n} 
\biggl( i\varepsilon_n + \frac{i\omega_\lambda}{2} \biggr)^\ell
({\mathcal G}_+)_a ({\mathcal G})_b ({\mathcal G})_c, \cr
&\tilde C^{(\ell)}_{cba} 
= -\lim_{\omega\rightarrow 0} \frac{k_{\rm B}T}{\hbar\omega} \sum_{n} 
\biggl( i\varepsilon_n + \frac{i\omega_\lambda}{2} \biggr)^\ell
({\mathcal G}_+)_c ({\mathcal G}_+)_b ({\mathcal G})_a, \cr
&D^{(\ell)}_{abcd} 
= -\lim_{\omega\rightarrow 0} \frac{k_{\rm B}T}{\hbar\omega} \sum_{n} 
\biggl( i\varepsilon_n + \frac{i\omega_\lambda}{2} \biggr)^\ell
({\mathcal G}_+)_a ({\mathcal G})_b ({\mathcal G})_c ({\mathcal G})_d, \cr
&\tilde D^{(\ell)}_{dcba} 
= -\lim_{\omega\rightarrow 0} \frac{k_{\rm B}T}{\hbar\omega} \sum_{n} 
\biggl( i\varepsilon_n + \frac{i\omega_\lambda}{2} \biggr)^\ell
({\mathcal G}_+)_d ({\mathcal G}_+)_c ({\mathcal G}_+)_b ({\mathcal G})_a,
\label{eq:DefDabcd}
\end{split}\end{equation}
where the substitution of $i\omega_\lambda \rightarrow \hbar(\omega+i\delta)$ 
should be made in the thermal Green's functions after taking the Matsubara summations.

The explicit forms of these functions are shown in Appendix C, where they are 
expressed in terms of the retarded and advanced Green's functions and the Fermi 
distribution function.
We can see that the relations
\begin{equation}
\left[ \tilde C^{(\ell)}_{cba} \right]^* = - C^{(\ell)}_{abc}, \qquad
\left[ \tilde D^{(\ell)}_{dcba} \right]^* = - D^{(\ell)}_{abcd},
\end{equation}
hold, which guarantee that the transport coefficients are real.

Because the matrix elements of $\gamma_\mu$ in Eq.~(\ref{eq:gammaDef}) 
have two terms, there are several types of contributions in $L^{(\ell)}_{\mu\nu}$
in Eq.~(\ref{eq:L11withD}) coming from different combinations of these two terms.
In particular, the contributions containing $D^{(\ell)}_{abcd}$ 
(i.e., the case with four Green's functions) are shown in Fig.~1. 
[Similar diagrams for $C_{abc}^{(\ell)}$ and $\tilde C_{abc}^{(\ell}$ can be drawn, which we do not show here.]
The diagrams in Fig.~1 are similar to those discussed in the case of the orbital 
magnetic susceptibility \cite{Ogata2015}.
As we will see later, part of the infinite summations is carried out by using the $f$-sum rule, which is indicated by the red squares in Fig.~1.

\def\MC#1{\textcolor{blue}{#1}}
\def\MCC#1{\textcolor{red}{#1}}
\begin{figure}[htbp]
\setlength{\unitlength}{1mm}
\centering\begin{picture}(80,140)
 \thicklines
  \put(7,130){\large $L_{\mu\nu}^{(\ell),1}$} \put(0,130){(a)} 
  \put(20,130){\line(1,0){7}} \put(22,125){$a$}  \put(30,130){\line(1,0){7}} \put(32,125){$a$}
  \put(40,130){\line(1,0){7}} \put(42,125){$a$}  \put(50,130){\line(1,0){7}} \put(52,125){$a$}  
  \put(27.25,129){\textcolor{blue}{$\otimes$}}  \put(37.24,129){\textcolor{blue}{$\otimes$}}
  \put(47.25,129){\textcolor{blue}{$\otimes$}}  \put(57.24,129){\textcolor{blue}{$\otimes$}}
  \put(26,125){\textcolor{blue}{$\frac{\partial \Eell}{\partial k_x}$}}
  \put(7,110){\large $L_{\mu\nu}^{(\ell),2}$} \put(0,115){(b)} 
  \put(20,110){\line(1,0){7}} \put(22,105){$a$}  \put(30,116){\line(1,0){7}} \put(32,111){$b$} 
  \put(40,110){\line(1,0){7}} \put(44,105){$a$}  \put(50,110){\line(1,0){7}} \put(52,105){$a$}  
  \MC{\multiput(27,110)(1,2){4}{\circle{1}} \multiput(40,110)(-1,2){4}{\circle{1}}
  \put(21.5,115){$p_{ab, \nu}$} \put(40.5,115){$p_{ba, x}$} }
  \put(47.25,109){\textcolor{blue}{$\otimes$}}  \put(57.24,109){\textcolor{blue}{$\otimes$}}

  \put(7,90){\large $L_{\mu\nu}^{(\ell),3}$} \put(0,95){(c)} 
  \put(20,90){\line(1,0){7}} \put(22,85){$a$}  \put(30,96){\line(1,0){7}} \put(32,91){$b$} 
  \put(40,96){\line(1,0){7}} \put(42,91){$b$}  \put(50,90){\line(1,0){7}} \put(52,85){$a$}  
  \MC{\multiput(27,90)(1,2){4}{\circle{1}} \multiput(50,90)(-1,2){4}{\circle{1}} \put(21.5,95){$p_{ab, \nu}$} \put(49.5,95){$p_{ba, y}$} }
  \put(37.24,95){\textcolor{blue}{$\otimes$}} \put(57.24,89){\textcolor{blue}{$\otimes$}}

  \put(7,70){\large $L_{\mu\nu}^{(\ell),4}$} \put(0,75){(d)} 
  \put(20,70){\line(1,0){7}} \put(22,65){$a$}  \put(30,76){\line(1,0){7}} \put(32,71){$b$} 
  \put(40,80){\line(1,0){7}} \put(42,75){$c$}  \put(50,70){\line(1,0){7}} \put(52,65){$a$}  
  \MC{\multiput(27,70)(1,2){4}{\circle{1}} \multiput(50,70)(-0.6,2){6}{\circle{1}}  \multiput(37,76)(1.8,2){3}{\circle{1}}
  \put(21.5,75){$p_{ab, \nu}$} \put(49.5,77){$p_{ca, y}$} \put(32,80){$p_{bc, x}$} } 
  \put(57.24,69){\textcolor{blue}{$\otimes$}}

  \put(7,50){\large $L_{\mu\nu}^{(\ell),5}$} \put(0,55){(e)} 
  \put(20,50){\line(1,0){7}} \put(22,46){$a$}  \put(30,56){\line(1,0){7}} \put(32,51){$b$} 
  \put(40,50){\line(1,0){7}} \put(45,46){$a$}  \put(50,56){\line(1,0){7}} \put(52,51){$b$}  

  \MC{\multiput(27,50)(1,2){4}{\circle{1}} \multiput(40,50)(-1,2){4}{\circle{1}} \put(21.5,55){$p_{ab, \nu}$} \put(39.5,55){$p_{ba, x}$}
  \multiput(47,50)(1,2){4}{\circle{1}} \multiput(60,50)(-1,2){4}{\circle{1}} \put(44,58){$p_{ab, y}$} \put(59.5,55){$p_{ba, \mu}$} }

  \put(7,25){\large $L_{\mu\nu}^{(\ell),6}$} \put(0,33){(f)} 
  \put(20,25){\line(1,0){7}} \put(22,21){$a$}  \put(30,31){\line(1,0){7}} \put(32,26){$b$} 
  \put(40,25){\line(1,0){7}} \put(45,21){$a$}  \put(50,35){\line(1,0){7}} \put(52,30){$c$}  

  \MC{\multiput(27,25)(1,2){4}{\circle{1}} \multiput(40,25)(-1,2){4}{\circle{1}} \put(21.5,30){$p_{ab, \nu}$} \put(39.5,30){$p_{ba, x}$}
  \multiput(47,25)(0.6,2){6}{\circle{1}} \multiput(60,25)(-0.6,2){6}{\circle{1}} \put(42,35){$p_{ac, y}$} \put(59,35){$p_{ca, \mu}$} }

  \put(7,2){\large $L_{\mu\nu}^{(\ell),7}$} \put(0,10){(g)} 
  \put(20,2){\line(1,0){7}} \put(22,-2){$a$}  \put(30,8){\line(1,0){7}} \put(32,3){$b$} 
  \put(50,5){\line(1,0){7}} \put(52,0){$d$}  \put(40,12){\line(1,0){7}} \put(42,7){$c$}  

  \MC{\multiput(27,2)(1,2){4}{\circle{1}} \multiput(37,8)(1.8,2){3}{\circle{1}} \put(21.5,7){$p_{ab. \nu}$} \put(32,12){$p_{bc. x}$} 
  \multiput(50,5)(-1,2){4}{\circle{1}} \multiput(60,2)(-1.5,1.5){3}{\circle{1}} \put(50,10){$p_{cd, y}$} \put(59,5){$p_{da, \mu}$} }

   \MCC{\put(24,105){\line(1,0){18}} \put(24,120){\line(1,0){18}} \put(24,105){\line(0,1){15}} \put(42,105){\line(0,1){15}}
            \put(24,45){\line(1,0){18}} \put(24,60){\line(1,0){18}} \put(24,45){\line(0,1){15}} \put(42,45){\line(0,1){15}} 
            \put(43,45){\line(1,0){18}} \put(43,60){\line(1,0){18}} \put(43,45){\line(0,1){15}} \put(61,45){\line(0,1){15}}
            \put(24,20){\line(1,0){18}} \put(24,38){\line(1,0){18}} \put(24,20){\line(0,1){18}} \put(42,20){\line(0,1){18}}  
            \put(43,20){\line(1,0){18}} \put(43,38){\line(1,0){18}} \put(43,20){\line(0,1){18}} \put(61,20){\line(0,1){18}} 
            \put(25,101.5){$f$-sum rule} \put(25,41.5){$f$-sum rule} \put(44,41.5){$f$-sum rule}
            \put(25,16.5){$f$-sum rule} \put(44,16.5){$f$-sum rule}
}
\end{picture}
\caption{Schematic representation of the contributions to $L_{\mu\nu}^{(\ell),1}$-$L_{\mu\nu}^{(\ell),7}$: 
The solid lines with band indices $a, b$ etc.\ represent 
the Green's functions. Height of these lines represents the energy level $\Eell$. 
The array of blue circles connecting the two lines represents the off-diagonal (or interband) matrix elements of 
$\gamma_\mu$, i.e., $p_{ab, \mu}$. 
The symbol $\otimes$ between the two solid lines represents the diagonal (or intraband) component of 
$\gamma_\mu$, i.e., $\partial \Eell/\partial k_\mu$.
The right-hand of each diagram is connected to its left-hand because of the trace in (\ref{PhiJJQ0}). 
The red squares indicate the part of the diagrams which can be expressed by 
the $f$-sum rule in Eq.~(\ref{fSumRule}).}
\label{Fig:01}
\end{figure}
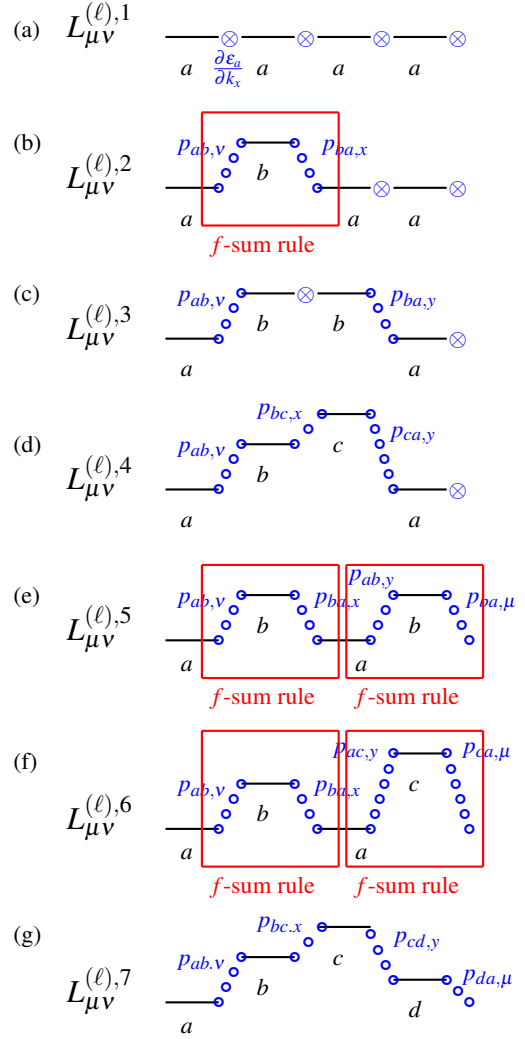

When we add all the diagrams in Fig.~1, we obtain the transport coefficients, 
which are valid for any strength of the scattering rate $\Gamma$. 
We noticed in Ref.~\cite{Konye2020} that, if we take the leading and subleading order 
contributions with respect to $\Gamma^\ell$, we can express $L^{(0)}_{\mu\nu}$
in terms of the Berry curvature and the orbital magnetic moment. 
In the following, we take a similar procedure to obtain $L^{(1)}_{\mu\nu}$ and
$L^{(2)}_{\mu\nu}$ in general.
Note that, in the case of the orbital magnetic susceptibility \cite{Ogata2015},
the contributions of the diagrams as in Fig.~1 can be summed. 
In that case, it was not necessary to introduce the relaxation rate $\Gamma$, 
because the susceptibility is a thermodynamic quantity. 
In other words, the calculation for the susceptibility was carried out in the order 
of $\Gamma^0$. 
In contrast, for the transport quantities, transport coefficients diverge for 
$\Gamma \rightarrow 0$ and we have to take into account the 
contributions proportional to $\Gamma^{-2}$ and $\Gamma^{-1}$, as we show below. 

First, we evaluate $C^{(\ell)}_{abc}$ and $D^{(\ell)}_{abcd}$ in the small $\Gamma$ expansion 
($\Gamma<<\varepsilon_{\rm F}$). Equations (\ref{eq:GexpforC1}), (\ref{eq:GexpforC2}) and
(\ref{eq:GexpforD}) in Appendix D show their contributions in the order of $\Gamma^{-2}$ and $\Gamma^{-1}$. 
As in usual transport theory, the contributions containing both the retarded Green's function $G_a^R$ and the advanced Green's function $G_a^A$ have singularities as $\Gamma^{-2}$ and $\Gamma^{-1}$.
Substituting the matrix elements in Eq.~(\ref{eq:gammaDef}) for $\gamma_\mu$ and the small $\Gamma$ expansions of $C^{(\ell)}_{abc}$ and $D^{(\ell)}_{abcd}$ into Eq.~(\ref{eq:L11withD}), we obtain each contribution $L_{\mu\nu}^{(\ell),1}$ to $L_{\mu\nu}^{(\ell),7}$ shown in Fig.~1. 
Their explicit expressions are shown in Eqs.~(\ref{eq:L1munu})-(\ref{eq:L6munu}) in Appendix D. 
[Note that $L^{(\ell),7}_{\mu\nu}$ does not appear since $D^{(\ell)}_{abcd}$ is 
in the order of $O(\Gamma^0)$.]
As a result, the transport coefficient can be obtained as
\begin{equation}
L^{(\ell)}_{\mu\nu} = \sum_{n=1}^2 L^{(\ell)C,n}_{\mu\nu} 
+ \sum_{n=1}^6 L^{(\ell),n}_{\mu\nu}
+O(\Gamma^0).
\label{eq:LmunuTotal}
\end{equation}

\subsection{Leading order of $O(\Gamma^{-2})$}

First, we consider the leading order terms in $L^{(\ell)}_{\mu\nu}$
in Eq.~(\ref{eq:LmunuTotal}), which are proportional to $\Gamma^{-2}$. 
As shown in Appendix D, they are $L^{(\ell)C,1}_{\mu\nu}$ in Eq.~(\ref{eq:LC1munu}), 
$L^{(\ell),1}_{\mu\nu}$ in Eq.~(\ref{eq:L1munu}), and the first term 
in $L^{(\ell),2}_{\mu\nu}$ in Eq.~(\ref{eq:L2munu}).
According to the different denominators in Eq.~(\ref{eq:L2munu}), 
we divide $L^{(\ell),2}_{\mu\nu}$ as
\begin{equation}
L^{(\ell),2}_{\mu\nu}=L^{(\ell),2;1}_{\mu\nu} 
+L^{(\ell),2;2}_{\mu\nu}+L^{(\ell),2;3}_{\mu\nu}.
\end{equation}
Each term has the denominator $\Gamma^2(\varepsilon_a-\varepsilon_b)$, 
$\Gamma(\varepsilon_a-\varepsilon_b)$, and $\Gamma(\varepsilon_a-\varepsilon_b)^2$,
respectively. 

As we can see from Fig.~1(a), $L_{\mu\nu}^{(\ell),1}$  
is purely intraband contribution, because only the intraband matrix 
elements of $\gamma_\mu$'s are involved. 
Considering that $\partial_x \varepsilon_a \partial [(\varepsilon_a-\mu)^\ell f'(\varepsilon_a)]/\partial \varepsilon_a = \partial_x [(\varepsilon_a-\mu)^\ell f'(\varepsilon_a)]$ 
and making integration by parts with respect to $k_x$ and $k_y$ symmetrically, we obtain
\begin{equation}\begin{split}
L_{\mu\nu}^{(\ell),1} &= -\frac{e^{3-\ell} B}{\hbar^2 V} \sum_{{\bm k},a} 
\frac{(\varepsilon-\mu)^\ell f'(\varepsilon_a)}
{8\Gamma^2} \bigl\{ 2 \Eellm \Eelln (\partial_x \partial_y \Eell) \cr
&+ \Eellm \Eellx (\partial_\nu \partial_y \Eell) 
 + \Eellm \Eelly (\partial_\nu \partial_x \Eell) \cr
&+ \Eelln \Eellx (\partial_\mu \partial_y \Eell) 
 + \Eelln \Eelly (\partial_\mu \partial_x \Eell) \bigr\}, 
\label{eq:L1munu2}
\end{split}\end{equation}
where $f(\varepsilon_a) = 1/(e^{\beta(\varepsilon_a-\mu)}+1)$ is the 
Fermi distribution function.  
In particular, when $\ell=0$ and $\mu=x, \nu=y$, i.e., 
in the case of Hall conductivity, this becomes
\begin{equation}\begin{split}
L_{xy}^{(0),1} &= -\frac{e^{3} B}{\hbar^2 V} \sum_{{\bm k},a} \frac{f'(\varepsilon_a)}{8\Gamma^2} \bigl\{ 4 \Eellx \Eelly (\partial_x \partial_y \Eell) \cr
&+ (\Eellx)^2  (\partial^2_y \Eell) + (\Eelly)^2 (\partial^2_x \Eell) \bigr\}, 
\label{Chi1}
\end{split}\end{equation}
which is similar to the result in the Boltzmann transport theory,
\begin{equation}\begin{split}
\sigma_{xy}^{\rm Boltzmann}&= -\frac{e^{3} B}{\hbar^2 V} \sum_{{\bm k},a} 
\frac{f'(\varepsilon_a)}
{8\Gamma^2} \bigl\{(\Eellx)^2 (\partial_y^2\Eell) \cr
&+ (\Eelly)^2 (\partial_x^2 \Eell) 
 -2\Eellx \Eelly (\partial_x \partial_y \Eell) \bigr\}, 
\label{eq:HallBoltzmann}
\end{split}\end{equation}
but the coefficient of the term $\Eellx \Eelly (\partial_x \partial_y \Eell)$ is different. 
We will show shortly that the result in the Boltzmann theory is obtained by adding other 
contributions from  $L_{\mu\nu}^{(0)C,1}$ and $L_{\mu\nu}^{(0),2:1}$ that are proportional to $1/\Gamma^2$.  
This means that one should not pick up only $L_{\mu\nu}^{(0),1}$ in discussing the 
Hall conductivity in a single-band model. 
This situation is the same as in the case of orbital magnetic susceptibility, where $\chi_1$ that 
looks like a purely intraband contribution gives a different result 
from the Landau-Peierls orbital magnetic susceptibility \cite{Fukuyama1971,Ogata2015}.

The first term of $L^{(\ell),2}_{\mu\nu}$ in Eq.~(\ref{eq:L2munu}) 
(i.e., $L^{(\ell),2;1}_{\mu\nu}$) 
has the summation over the band indices $a$ and $b$ ($b\ne a$) [see also Fig.~1(b)]. 
Therefore, one may think that this term is an interband contribution. 
However, the summation over $b$ can be carried out exactly, and then this term can be 
regarded as an intraband contribution, which should be added to Eq.~(\ref{eq:L1munu2}). 
For this purpose, we use the generalized $f$-sum rule,
\begin{equation}
\sum_{b, (b \ne a)} \frac{p_{ab, \mu} p_{ba, \nu}}{\Eell-\varepsilon_b} 
= \frac{1}{2}\left(
\partial_\mu \partial_\nu \Eell 
-\frac{\hbar^2}{m} \delta_{\mu\nu} \right) + i\ M_{\mu\nu}^a,
\label{fSumRule}
\end{equation}
where $M_{\mu\nu}^a$ is the orbital magnetic moment for the band $a$ defined before in Eq.~(\ref{eq:MmunuDef}).
Proof of this identity is given in Appendix E. 
When we use this generalized $f$-sum rule, we obtain
\begin{equation}\begin{split}
L^{(\ell),2;1}_{\mu\nu} &= \frac{e^{3-\ell} B}{\hbar^2 V} \sum_{{\bm k},a}
\frac{(\varepsilon_a-\mu)^\ell f'(\varepsilon_a)}{4\Gamma^2} \cr
&\biggl[ \Eelln \Eellx \bigg(\partial_y \partial_\mu \Eell - \frac{\hbar^2}{m} \delta_{\mu y} \bigg)
+ \Eellm \Eelln (\partial_x \partial_y \Eell) \cr
&+\Eellm \Eelly \left(\partial_\nu \partial_x \Eell - \frac{\hbar^2}{m} \delta_{\nu x} \right) \biggr].
\label{eq:L2munu1}
\end{split}\end{equation}
This $f$-sum rule is schematically shown in Fig.~\ref{Fig:01}(b), 
in which the red square indicates the part of the diagram representing the $f$-sum rule.  

Thus, the total of Eqs.~(\ref{eq:LC1munu}), (\ref{eq:L1munu2}), and (\ref{eq:L2munu1}) becomes
\begin{equation}\begin{split}
L^{(\ell),C1}_{\mu\nu} &+ L^{(\ell),1}_{\mu\nu} + L^{(\ell),2;1}_{\mu\nu}=
-\frac{e^{3-\ell} B}{\hbar^2 V} \sum_{{\bm k},a} 
\frac{(\varepsilon_a-\mu)^\ell f'(\varepsilon_a)}{8\Gamma^2} \cr
&\times \bigl\{ \Eellm \Eellx (\partial_\nu \partial_y \Eell) 
 - \Eellm \Eelly (\partial_\nu \partial_x \Eell) \cr
&- \Eelln \Eellx (\partial_\mu \partial_y \Eell) 
 + \Eelln \Eelly (\partial_\mu \partial_x \Eell) \bigr\}. 
\label{eq:sigma1c}
\end{split}\end{equation}
In particular, when $\ell=0$, $\mu=x$, and $\nu=y$, i.e., in the case of Hall 
conductivity, this total becomes equal to the result obtained using Boltzmann theory. 

The generalized $f$-sum rule in Eq.~(\ref{fSumRule}) results from the completeness 
property of $\Uellb$ (see Appendix A of Ref.~\cite{Ogata2015}).  
One may call the left-hand side of (\ref{fSumRule}) as \lq\lq interband" since it 
contains the off-diagonal matrix element $p_{ab, \mu}$. 
On the other hand, the right-hand side of (\ref{fSumRule}) is expressed by a 
single-band property, $\Eell$, 
and as a result, $L^{(\ell),2;1}_{\mu\nu}$ looks like an 
\lq\lq intraband" contribution.  
This indicates that the naive classification of \lq\lq intraband" and \lq\lq interband" 
does not apply, as discussed for the case of orbital magnetic susceptibility \cite{Ogata2015}.

\subsection{Subleading order of $O(\Gamma^{-1})$}

Next, we discuss the subleading order of $O(\Gamma^{-1})$. 
The second term of $L^{(\ell),2}_{\mu\nu}$ in Eq.~(\ref{eq:L2munu}) 
(i.e., $L^{(\ell),2;2}_{\mu\nu}$) can be calculated in a similar way to 
$L^{(\ell),2;1}_{\mu\nu}$, which yields
\begin{equation}\begin{split}
L^{(\ell),2;2}_{\mu\nu} &= \frac{e^{3-\ell} B}{\hbar^2 V} \sum_{{\bm k},a}
\frac{\frac{\partial}{\partial\varepsilon_a}
[(\varepsilon_a-\mu)^\ell f'(\varepsilon_a)]}{2\Gamma}
\biggl[ \Eelln \Eellx M_{y\mu}^a \cr
&+ \Eellm \Eelln M_{xy}^a +\Eellm \Eelly M_{\nu x}^a \biggr].
\label{eq:L2munu2}
\end{split}\end{equation}

The third term of $L^{(\ell),2}_{\mu\nu}$ 
(i.e., $L^{(\ell),2;3}_{\mu\nu}$) has the denominator $(\varepsilon_a-\varepsilon_b)^2$
that is different from $L^{(\ell),2;1}_{\mu\nu}$ or $L^{(\ell),2;2}_{\mu\nu}$ 
discussed above. 
In this case, instead of Eq.~(\ref{fSumRule}), we use another generalized $f$-sum rule 
(see Appendix E)
\begin{equation}\begin{split}
\sum_{b, (b \ne a)} \frac{p_{ab, \mu} p_{ba, \nu}}{(\Eell-\Eellb)^2} 
&={\rm Re} \langle \partial_\mu a | \partial_\nu a \rangle 
- \langle \partial_\mu a | a \rangle \langle a | \partial_\nu a \rangle
-\frac{i}{2} \ \Omega_{\mu\nu}^a,
\label{fSumRule2}
\end{split}\end{equation}
where $\Omega_{\mu\nu}^a$ is the Berry curvature for the band $a$ defined before in Eq.~(\ref{eq:OmegamunuDef}).

Using this second generalized $f$-sum rule, we obtain
\begin{equation}\begin{split}
L^{(\ell),2;3}_{\mu\nu} &= \frac{e^{3-\ell} B}{\hbar^2 V} \sum_{{\bm k},a}
\frac{(\varepsilon_a-\mu)^\ell f'(\varepsilon_a)}{2\Gamma}
\biggl[ \Eelln \Eellx \Omega_{y\mu}^a \cr
&+ \Eellm \Eelln \Omega_{xy}^a +\Eellm \Eelly \Omega_{\nu x}^a \biggr].
\label{eq:L2munu3}
\end{split}\end{equation}
Note that we have used the fact that 
$\langle \partial_\mu a | b \rangle = -\langle a | \partial_\mu b \rangle$
and it is purely imaginary for any $\mu$ when $a$ is equal to $b$.

Features similar to $L^{(\ell),2}_{\mu\nu}$ are present in 
$L^{(\ell),5}_{\mu\nu}$ and $L^{(\ell),6}_{\mu\nu}$, as seen in 
Fig.\ref{Fig:01}, where the excluded term of $b=c$ in $L^{(\ell),6}_{\mu\nu}$ is 
supplemented by $L^{(\ell),5}_{\mu\nu}$, leading to the independent summations over  
$b$ and $c$.
Other terms are calculated similarly, whose details are shown in Appendix F. 

Combining the leading and subleading order terms we obtain the total thermoelectric conductivities as
shown in Eq.~(\ref{eq:LmunuFinal}).
Note that this final formula does not contain the terms proportional to $\hbar^2/m$ which appeared in Eq.~(\ref{eq:LmunuTotal}) as $L^{(\ell)C,1}_{\mu\nu}$ and $L^{(\ell)C,2}_{\mu\nu}$. 
For example, for the Hall conductivity
\begin{equation}\begin{split}
L_{xy}^{(\ell)} &= -\frac{e^{3-\ell} B}{\hbar^2 V}\sum_{{\bm k},a} 
(\varepsilon_a-\mu)^\ell f'(\varepsilon_a)\cr
\times &\biggl[ \frac{1}{8\Gamma^2}
\bigl\{ (\Eellx)^2 (\partial_y^2 \Eell)  + (\Eelly)^2 (\partial_x^2 \Eell) \cr
&- 2\Eellx \Eelly (\partial_x \partial_y \Eell) \bigr\} 
+\frac{1}{2\Gamma} \Eellx \Eelly \Omega_{xy}^a \cr
&-\frac{1}{4\Gamma} \bigl\{ \Eellx \partial_y M_{xy} 
+ \Eelly \partial_x M_{xy} -2(\partial_x \partial_y \Eell) M_{xy} \bigr\} \biggr].
\end{split}\end{equation}
and for the $xx$- and $zz$-component,
\begin{equation}\begin{split}
L_{xx}^{(\ell)} &= -\frac{e^{3-\ell} B}{\hbar^2 V}\sum_{{\bm k},a} 
(\varepsilon_a-\mu)^\ell f'(\varepsilon_a)\cr
\times &\biggl[ \frac{1}{2\Gamma} (\partial_x \Eell)^2 \Omega_{xy}^a 
-\frac{1}{2\Gamma} \bigl\{ \partial_x \Eell \partial_x M_{xy} 
- (\partial_x^2 \Eell) M_{xy} \bigr\} \biggr].\cr
L_{zz}^{(\ell)} &= -\frac{e^{3-\ell} B}{\hbar^2 V}\sum_{{\bm k},a} 
(\varepsilon_a-\mu)^\ell f'(\varepsilon_a)\cr
\times &\biggl[ -\frac{1}{2\Gamma} \bigl\{ (\partial_z \Eell)^2 \Omega_{xy}^a 
 + 2\Eellx \partial_z \Eell \Omega_{yz}^a +2 \Eelly \partial_z \Eell \Omega_{zx}^a \bigr\} \cr
&-\frac{1}{2\Gamma} \bigl\{ \partial_z \Eell \partial_z M_{xy} 
- (\partial_z^2 \Eell) M_{xy} \bigr\} \biggr].
\end{split}\end{equation}
These results are consistent with our previous result for $\ell=0$ obtained in Ref.~\cite{Konye2020}.

\section{Linear magnetothermopower}
In this section we study a simple two-band model that shows non-trivial effects, such as linear magnetoconductivity and magnetothermopower. We consider the following gapped model with broken time-reversal symmetry:
\begin{equation}
\label{eq:Ham}
	H = v \hbar k_x\sigma_x+v \hbar k_y\sigma_y+\Delta(1+\delta k_z^2)\sigma_z.
\end{equation}
The dispersion relation of this model is
\begin{equation}
    E_\pm = \pm \Delta \sqrt{\frac{v^2\hbar^2}{\Delta^2}\left(k_x^2+k_y^2\right)+(1+\delta k_z^2)^2}.
\end{equation}

In order to observe non-trivial effects that violate the Onsager relations, time-reversal symmetry must be broken. In the case of completely gapless linear spectrum the conductivity will be chemical potential independent \cite{Konye2020} this leads to a vanishing thermopower. The presence of the gap leads to a finite thermopower caused by the diverging density of states around the gap.

The Eq.~\eqref{eq:Ham} system has a non-vanishing Berry curvature and orbital magnetic moment expressed as
\begin{align}
    \Omega_{xy}^{\pm} &= - \hbar^2 v^2 \Delta \frac{1 + \delta k_z^2}{2E_\pm^3}, &
    \Omega_{yz}^{\pm} &= - \hbar^2 v^2 \Delta \frac{\delta k_xk_z}{E_\pm^3},\\
    \Omega_{zx}^{\pm} &= - \hbar^2 v^2 \Delta\frac{\delta k_yk_z}{E_\pm^3}, &
    M_{xy}^{\pm} &= \hbar^2 v^2 \Delta\frac{1+\delta k_z^2}{2E_\pm^2}.
\end{align}

Considering the symmetries of the system the magnetoconductivity is of the form
\begin{equation}
    \sigma = \begin{pmatrix}
    \sigma_{xx} & \sigma_{xy} & 0\\
    -\sigma_{xy} & \sigma_{xx} & 0\\
    0 & 0 & \sigma_{zz}\\
    \end{pmatrix}.
\end{equation}

There are three independent components. In the linear order of the magnetic field these can be divided as:
\begin{subequations}
\begin{align}
    \sigma_{xx} &= \sigma_{xx}^{(0)}+\sigma_{xx}^{(1;1)},\\
    \sigma_{zz} &= \sigma_{zz}^{(0)}+\sigma_{zz}^{(1;1)},\\
    \sigma_{xy} &= \sigma_{xy}^{(1;0)},
\end{align}
\end{subequations}
where $(0)$ is the $B=0$ result, $(1;0)$ is the $B$ linear term with $\Gamma^{-2}$ and $(1;1)$ with $\Gamma^{-1}$.

\begin{figure}
	\includegraphics[width=\columnwidth]{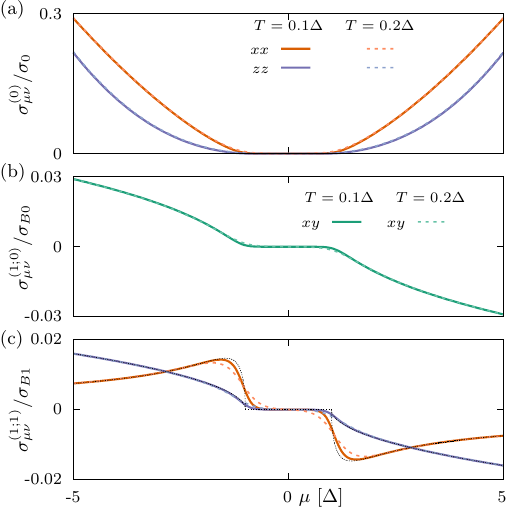}
\caption{\label{fig:cond}Different components of the conductivity tensor of Eq.~\eqref{eq:Ham} as a function of chemical potential up to linear order of the magnetic field. (a) Conductivity for $B=0$ in units of $\sigma_0=e^2\Delta^2/\Gamma v\hbar^2$. (b) Hall-conductivity in units of $\sigma_{B0}=e^3B\Delta v/\Gamma^2\hbar$. (c) Longitudinal and transverse linear magnetoconductivities for 3 different temperatures in units of $\sigma_{B1}=e^3B v/\Gamma\hbar$. In all panels $\delta = 0.1 (\hbar v)^2/\Delta^2$ and only the non-zero components are shown.}
\end{figure}

The $B=0$ result can be computed using Boltzmann theory (discarding $\mathcal{O}(\Gamma^0)$ terms). 
The rest of the terms can be computed using Eq.~\eqref{eq:LmunuFinal}. All terms are computed numerically, using Monte-Carlo integration.
The results as a function of the chemical potential are shown in Fig.~\ref{fig:cond}. Panels (a) and (b) show the normal conductivity with symmetric diagonal elements and antisymmetric Hall conductivity. Panel (c) shows the linear magnetoconductivity terms that are classically forbidden by the Onsager relations where the conductivity should satisfy $\sigma_{\mu\nu}(\vb{B})=\sigma_{\nu\mu}(-\vb{B})$.

\begin{figure}
	\includegraphics[width=\columnwidth]{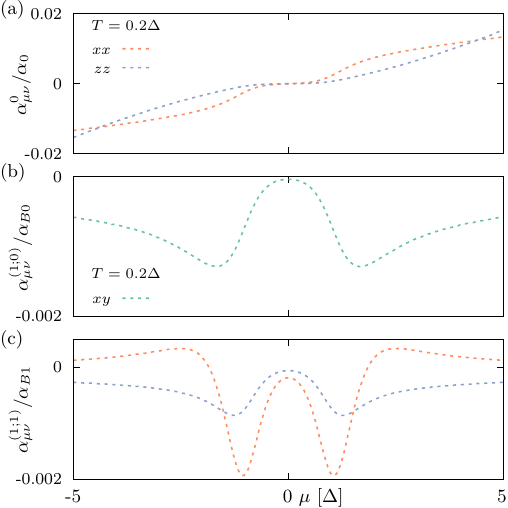}
\caption{\label{fig:thermel}Different components of the thermoelectric conductivity tensor of Eq.~\eqref{eq:Ham} as a function of chemical potential up to linear order of the magnetic field. All panels correspond to the panels in Fig.~\ref{fig:cond}, with the same parameters. The thermoelectric conductivity is measured in units $\alpha_{0/B0/B1}=(\Delta/e)\sigma_{0/B0/B1}$}
\end{figure}

Similarly to the conductivity the thermoelectric conductivity ($\alpha = \vb{L}_{12}$) will be of the form
\begin{equation}
    \alpha = \begin{pmatrix}
    \alpha_{xx} & \alpha_{xy} & 0\\
    -\alpha_{xy} & \alpha_{xx} & 0\\
    0 & 0 & \alpha_{zz}\\
    \end{pmatrix},
\end{equation}
with
\begin{subequations}
\begin{align}
    \alpha_{xx} &= \alpha_{xx}^{(0)}+\alpha_{xx}^{(1;1)},\\
    \alpha_{zz} &= \alpha_{zz}^{(0)}+\alpha_{zz}^{(1;1)},\\
    \alpha_{xy} &= \alpha_{xy}^{(1;0)}.
\end{align}
\end{subequations}
This is computed similarly to the conductivity and the results as a function of the chemical potential are shown in Fig.~\ref{fig:thermel}.
The results for the thermoelectric conductivity also show finite $\alpha_{\mu\mu}^{(1;1)}$ terms that are classically forbidden, these appear because of the non-constant chemical potential dependence of the $\sigma_{\mu\mu}^{(1;1)}$ terms. We see that if the conductivity is an even (odd) function of the chemical potential the thermoelectric conductivity is odd (even). This is related to the charge carriers being electrons or holes with opposite charge.

\begin{figure}
	\includegraphics[width=\columnwidth]{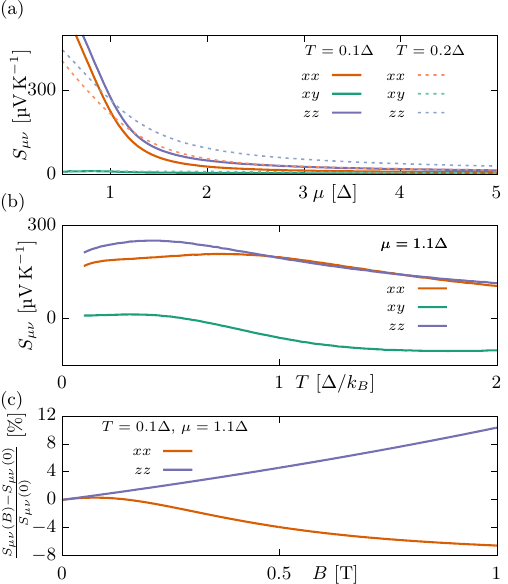}
\caption{\label{fig:seeb}Different components of the Seebeck tensor of \eqref{eq:Ham} up to linear order of the magnetic field as a function of (a) chemical potential at $B=\SI{1}{\tesla}$, (b) temperature at $B=\SI{1}{\tesla}$, and (c) magnetic field. The parameters used are $v=10^6\si{\meter\per\second}$, $\Delta=\SI{0.1}{\electronvolt}$, $\Gamma=0.01\Delta$, $\delta = 0.1 (\hbar v)^2/\Delta^2$}
\end{figure}

Using Eq. \eqref{eq:transport} we can obtain the Seebeck coefficient tensor from the conductivity and thermoelectric conductivity, which will have the same structure as the former two with
\begin{subequations}
\begin{align}
S_{xx} &= \frac{1}{T}\frac{\alpha_{xx}\sigma_{xx}+\alpha_{xy}\sigma_{xy}}{\sigma_{xx}^2+\sigma_{xy}^2},\\
S_{zz} &= \frac{1}{T}\frac{\alpha_{zz}}{\sigma_{zz}},\\
S_{xy} &= \frac{1}{T}\frac{\alpha_{xx}\sigma_{xy}-\alpha_{xy}\sigma_{xx}}{\sigma_{xx}^2+\sigma_{xy}^2}.
\end{align}
\end{subequations}
Using the previously computed results and choosing realistic parameters ($v=10^6\si{\meter\per\second}$, $\Delta=\SI{0.1}{\electronvolt}$, $\Gamma=0.01\Delta$) the thermopower is calculated and shown in Fig.~\ref{fig:seeb} as a function of chemical potential, temperature and magnetic field.

We can see that the Seebeck tensor gets enhanced close to the band gap.
We get a finite but small Nernst coefficient $S_{xy}$.

If the $B$-linear terms such as $\sigma^{(1;1)}$ and $\alpha^{(1;1)}$ are zero $S_{xx}(B)-S_{xx}(0)\propto B^2$ and $S_{zz}=\text{const.}$ If these non-trivial terms exist the magnetothermopower can have a linear contribution in both cases. 
In the case of $S_{zz}$ we can see this linear dependence in Fig.~\ref{fig:seeb}(c).
For $S_{xx}$ we get a more complicated magnetic field dependence.

\begin{figure}
	\includegraphics[width=\columnwidth]{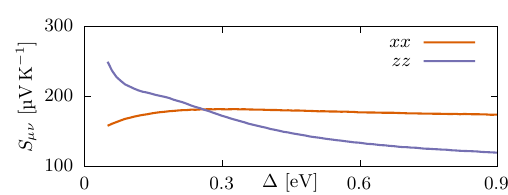}
\caption{\label{fig:Sgap} Seebeck tensor as a function of the gap $\Delta$. The parameters used are $v=10^6\si{\meter\per\second}$, $\Gamma=\SI{0.001}{\electronvolt}$, $\mu = \Delta+\SI{0.01}{\electronvolt}$, $k_BT=\SI{0.1}{\electronvolt}$, $\delta = 10 (\hbar v)^2/\si{\square\electronvolt}$}
\end{figure}

To see the gap dependence we computed the Seebeck coefficients as a function of $\Delta$ as shown in Fig.~\ref{fig:Sgap}.
We can see that the Seebeck coefficient in the $x$ direction is less sensitive to the gap size, while in the $z$ direction it is strongly enhanced for small gaps.

\begin{figure}
	\includegraphics[width=\columnwidth]{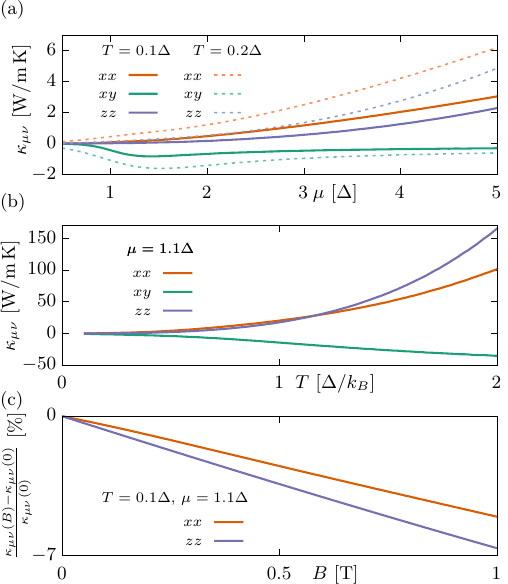}
\caption{\label{fig:thermocond}Different components of the thermal conductivity tensor of \eqref{eq:Ham} up to linear order of the magnetic field. The panels and parameters are the same as in Fig.~\ref{fig:seeb}}
\end{figure}

\begin{figure}
	\includegraphics[width=\columnwidth]{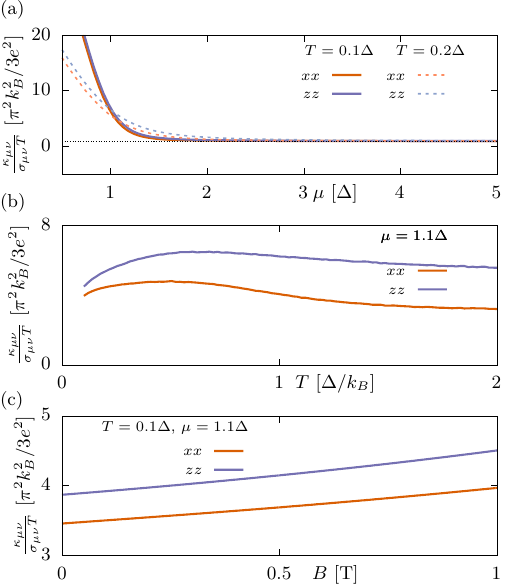}
\caption{\label{fig:Lorentz}The Lorenz number in the $x$ and $z$ directions of \eqref{eq:Ham} up to linear order of the magnetic field. The panels and parameters are the same as in Fig.~\ref{fig:seeb}}
\end{figure}

We also compute the thermal conductivity (see Fig.~\ref{fig:thermocond}) and the ratio of thermal conductivity to electric conductivity also known as Lorentz number (see Fig.~\ref{fig:Lorentz}).
Far away from the gap at large $\mu$ we recover the Wiedemann-Franz law with $L=\kappa/\sigma T = \pi^2k_B^2/3e^2$.
Closer to the band gap the Lorentz number gets enhanced and takes large values.
As a function of magnetic field (see Fig.~\ref{fig:Lorentz}c) we get a weak linear increase.

\begin{figure}
	\includegraphics[width=\columnwidth]{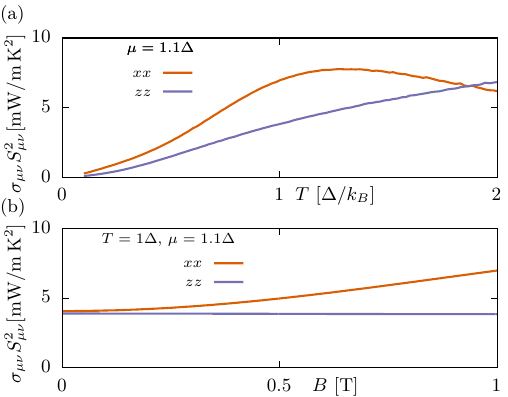}
\caption{\label{fig:power} Power factor of \eqref{eq:Ham} up to linear order of the magnetic field. The panels and parameters are the same as in Fig.~\ref{fig:seeb}}
\end{figure}

\begin{figure}
	\includegraphics[width=\columnwidth]{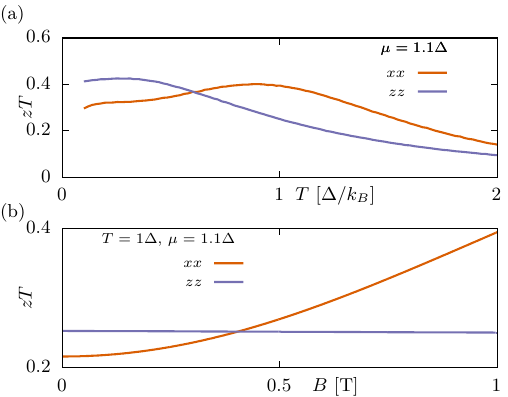}
\caption{\label{fig:ztfactor} Figure of merit ($zT$) of \eqref{eq:Ham} up to linear order of the magnetic field. The panels and parameters are the same as in Fig.~\ref{fig:seeb}}
\end{figure}

Finally, we compute the power factor defined as $\sigma S^2$ (see Fig.~\ref{fig:power}) and the figure of merit $zT=\sigma S^2 T/\kappa$ (see Fig.~\ref{fig:ztfactor}).
These are experimentally relevant quantities characterizing good thermoelectric materials.
A high figure of merit means more efficient conversion of heat energy into electrical energy.
The results of the power factor and figure of merit show a strong enhancement as a function of the magnetic field in the $x$ direction.
However, note that the contribution of phonons to the thermal conductivity, $\kappa$, 
is not taken into account, so the figure of merit in Fig.~\ref{fig:ztfactor} is 
generally suppressed by the presence of the phonon thermal conductivity.

\section{Conclusion}

In this paper we presented an analytic formalism to calculate thermoelectric coefficients in the presence of a low magnetic field up to linear order in the magnetic field.
The formulas are based on linear response theory and are taken in the limit of small scattering rates.
The resulting expression for the transport coefficients in Eq.~\eqref{eq:LmunuFinal} is general and can be applied to any system described by a single particle Bloch Hamiltonian or effective Bloch Hamiltonian.

In order to showcase the non-trivial effects that can be studied with our formalism we computed different magnetothermoelectric observable quantities for a simple model of an insulator with broken time-reversal symmetry.
Such an effect is a linear longitudinal magnetothermopower related to the presence of non-vanishing Berry curvature and orbital magnetic moment.

\bigskip\noindent
{\bf Acknowledgment}

We thank very fruitful discussions with H.\ Fukuyama, H.\ Matsuura, and H.\ Maebashi. 
This work was supported by a Grant-in-Aid for Scientific Research (Grants No.
JP23H01118 and JP23K03274), and JST-Mirai Program Grant Number JPMJMI19A1.

\appendix

\section{Current and heat current operators in the presence of magnetic field}

In the presence of a magnetic field, we use the Hamiltonian
\begin{equation}
\begin{split}
        H({\bm r}) &=
\frac{1}{2m}({\bm p}-e{\bm A}({\bm r}))^2 
+ V({\bm r}) +\frac{\hbar^2}{8m^2c^2} {\bm \nabla}^2 V({\bm r}) \cr
&- \frac{e\hbar}{2m} {\bm \sigma}\cdot \left\{ {\bm b}({\bm r}) + {\bm B}({\bm r}) \right\}
+ \frac{\hbar}{4m^2c^2} {\bm \sigma}\cdot {\bm \nabla} V \times ({\bm p} - e{\bm A}({\bm r})),
\end{split}
\end{equation}
with ${\bm B}({\bm r}) = {\rm rot}\ {\bm A}({\bm r})$, 
which is an extension of Eq.~(\ref{Hamiltonian}).

We derive the current operator using the continuity equation of Eq.~(\ref{eq:continuity}). 
The time derivative of the charge density is 
\begin{equation}
\frac{d}{dt} e\rho({\bm r}) = -\frac{ie}{\hbar} [\rho ({\bm r}), {\mathcal H}]
= -\frac{ie}{\hbar} \psi^\dagger(\bm r) \{ H({\bm r})-\overleftarrow H ({\bm r}) \} \psi(\bm r).
\end{equation}
Then, we can show that
\begin{equation}
H({\bm r})-\overleftarrow H ({\bm r}) = 
({\bm p} - \overleftarrow {\bm p}) \overleftrightarrow{\bm j_{0A}}(\bm r)
=-i\hbar({\bm \nabla}+\overleftarrow{\bm \nabla})\overleftrightarrow{\bm j_{0A}}(\bm r),
\label{eq:AppendixA1}
\end{equation}
where $\overleftrightarrow{\bm j_{0A}}(\bm r)$ is defined as
\begin{equation}
\overleftrightarrow {{\bm j}_{0A}}({\bm r})
=\frac{1}{2m} (-i\hbar {\bm \nabla} + i\hbar \overleftarrow{\bm \nabla} -2e{\bm A})
+ \frac{\hbar}{4m^2c^2} {\bm \sigma}\times {\bm \nabla}V.
\label{eq:jdefsA}
\end{equation}
Therefore, we obtain
\begin{equation}
\frac{d}{dt} e\rho({\bm r}) 
= -e {\bm \nabla} \left[ \psi^\dagger(\bm r) \overleftrightarrow{\bm j_{0A}}(\bm r) \psi(\bm r)
\right],
\end{equation}
which leads to the current density operator
\begin{equation}
\bm j(\bm r)=e\psi^\dagger(\bm r)\overleftrightarrow{\bm j_{0A}}(\bm r)\psi(\bm r).
\end{equation}
The contribution proportional to the vector potential $\bm A$ is the diamagnetic current.
When ${\bm A}={\bm 0}$, this current density operator becomes the one in Eq.~(\ref{eq:Jtot}).

Similarly, the time derivative of the energy density is 
\begin{equation}\begin{split}
\frac{d}{dt} h({\bm r}) &= -\frac{i}{\hbar} [h({\bm r}), {\mathcal H}] \cr
&= -\frac{i}{2\hbar} \psi^\dagger(\bm r) \bigl[ \{ H({\bm r})-\overleftarrow H ({\bm r}) \} H({\bm r})
\cr
&+ \overleftarrow H ({\bm r}) \{ H({\bm r})-\overleftarrow H ({\bm r}) \} \bigr] \psi(\bm r).
\end{split}\end{equation}
Again using Eq.~(\ref{eq:AppendixA1}), we obtain
\begin{equation}
\frac{d}{dt} h({\bm r}) 
= -{\bm \nabla} \left[ \frac{1}{2} \psi^\dagger(\bm r) 
\left\{ \overleftrightarrow{\bm j_{0A}}(\bm r) H({\bm r})
+ \overleftarrow H ({\bm r}) \overleftrightarrow{\bm j_{0A}}(\bm r) \right\} \psi(\bm r)\right],
\end{equation}
which leads to the energy current operator 
\begin{equation}
\bm j_{E} (\bm r)= \frac{1}{2} \psi^\dagger(\bm r) \left\{ 
\overleftrightarrow{\bm j_{0A}}(\bm r)  H (\bm r)
+ \overleftarrow H (\bm r) \overleftrightarrow{\bm j_{0A}}(\bm r)\right\} \psi(\bm r).
\end{equation}

Finally, let us derive the relationship between the current and heat current operator in 
Eq.~(\ref{JMrelation}).
Using the imaginary time interaction representation, we obtain
\begin{equation}\begin{split}
&\frac{1}{2e} \left( \frac{\partial}{\partial \tau} - \frac{\partial}{\partial \tau'} \right)
{\bm j} ({\bm r},\tau,\tau') \cr
&= \frac{1}{2} \bigl\{ [{\mathcal H} -\mu N, \psi^\dagger(\bm r, \tau)] 
\overleftrightarrow{\bm j_{0A}}(\bm r) \psi(\bm r, \tau') \cr
&- \psi^\dagger(\bm r) \overleftrightarrow{\bm j_{0A}}(\bm r, \tau) 
[{\mathcal H} - \mu N, \psi(\bm r, \tau')] \bigr\}\cr
&= \frac{1}{2} \bigl\{ \psi^\dagger(\bm r, \tau) (\overleftarrow H({\bm r}) -\mu )
\overleftrightarrow{\bm j_{0A}}(\bm r) \psi(\bm r, \tau') \cr
&+ \psi^\dagger(\bm r) \overleftrightarrow{\bm j_{0A}}(\bm r, \tau) 
( H({\bm r}) -\mu ) \psi(\bm r, \tau') \bigr\} \cr
&= {\bm j}_E ({\bm r},\tau,\tau') -\mu {\bm j} ({\bm r},\tau,\tau')/e.
\end{split}\end{equation}

\section{Generalization of Fukuyama's formula of Hall conductivity}

Fukuyama \cite{Fukuyama1969} used the Luttinger-Kohn representation of the Bloch 
wave functions and obtained the current-current correlation function $\Phi_{xy}(i\omega_\lambda)$ in the first order of the magnetic field. 
That formula can be generalized for the cases with other components ($\mu, \nu = x,y,z$) as follows:
\begin{equation}\begin{split}
&\Phi_{\mu\nu}^{11} (i\omega_\lambda) =\cr
& \frac{k_{\rm B}T}{V} \sum_{n,{\bm k}} 
\frac{i|e|^3 B}{2m\hbar} {\rm Tr}\bigl[ 
-\delta_{\nu y} (\gamma_\mu {\mathcal G}_+ {\mathcal G}\gamma_x {\mathcal G} 
-\gamma_\mu {\mathcal G}_+ \gamma_x {\mathcal G}_+ {\mathcal G}) \cr
&\hskip 20 truemm +\delta_{\nu x} (
\gamma_\mu {\mathcal G}_+ {\mathcal G} \gamma_y {\mathcal G}
-\gamma_\mu {\mathcal G}_+ \gamma_y {\mathcal G}_+ {\mathcal G})
 \bigr] \cr
&+ \frac{k_{\rm B}T}{V} \sum_{n,{\bm k}}\frac{i|e|^3 B}{2\hbar^3} {\rm Tr}\ \bigl[  
\gamma_\mu {\mathcal G}_+ \gamma_x {\mathcal G}_+ \gamma_\nu {\mathcal G} \gamma_y {\mathcal G} 
-\gamma_\mu {\mathcal G}_+ \gamma_y {\mathcal G}_+ \gamma_\nu {\mathcal G} \gamma_x {\mathcal G} \cr
&\hskip 20 truemm 
+\gamma_\mu {\mathcal G}_+ \gamma_\nu {\mathcal G} \gamma_x {\mathcal G} \gamma_y {\mathcal G} 
-\gamma_\mu {\mathcal G}_+ \gamma_\nu {\mathcal G} \gamma_y {\mathcal G} \gamma_x {\mathcal G} \cr
&\hskip 20 truemm 
+\gamma_\mu {\mathcal G}_+ \gamma_x {\mathcal G}_+ \gamma_y {\mathcal G}_+ \gamma_\nu {\mathcal G} 
-\gamma_\mu {\mathcal G}_+ \gamma_y {\mathcal G}_+ \gamma_x {\mathcal G}_+ \gamma_\nu {\mathcal G} \bigr],
\label{HFfirst}
\end{split}\end{equation}
where the magnetic field is parallel to the $z$-axis. 
In the Luttinger-Kohn representation, the relations 
\begin{equation}
\frac{\partial}{\partial k_\mu} \mathcal G = \mathcal G \gamma_\mu \mathcal G, 
\qquad \frac{\partial}{\partial k_\nu}\gamma_\mu = \frac{\hbar^2}{m} \bm 1 \delta_{\mu\nu},
\end{equation}
hold, where $\bm 1$ represents the unit matrix. 
Therefore, using these relations and integration by parts, we can 
rewrite the first term in the second summation in Eq.~(\ref{HFfirst}) as
\begin{equation}\begin{split}
&\sum_{\bm k} \gamma_\mu {\mathcal G}_+ \gamma_x {\mathcal G}_+ \gamma_\nu {\mathcal G} \gamma_y {\mathcal G} 
=\sum_{\bm k} \gamma_\mu {\mathcal G}_+ \gamma_x {\mathcal G}_+ \gamma_\nu \frac{\partial}{\partial k_y} {\mathcal G} \cr
=&-\sum_{\bm k} \frac{\hbar^2}{m} \left\{ \delta_{\mu y} {\mathcal G}_+ \gamma_x {\mathcal G}_+ \gamma_\nu {\mathcal G} 
+\delta_{\nu y} \gamma_\mu {\mathcal G}_+ \gamma_x {\mathcal G}_+ {\mathcal G} \right\} \cr
&-\sum_{\bm k} \left\{ 
 \gamma_\mu {\mathcal G}_+ \gamma_y {\mathcal G}_+ \gamma_x {\mathcal G}_+ \gamma_\nu {\mathcal G} 
+\gamma_\mu {\mathcal G}_+ \gamma_x {\mathcal G}_+ \gamma_y {\mathcal G}_+ \gamma_\nu {\mathcal G} 
\right\},
\end{split}\end{equation}
for any $\mu$ and $\nu$. 
A similar integration by parts for the second term in the second summation in
Eq.~(\ref{HFfirst}) 
\begin{equation}
-\sum_{\bm k} \gamma_\mu {\mathcal G}_+ \gamma_y {\mathcal G}_+ \gamma_\nu {\mathcal G} \gamma_x {\mathcal G} 
=-\sum_{\bm k} \gamma_\mu \left( \frac{\partial}{\partial k_y} {\mathcal G}_+ \right) 
\gamma_\nu  {\mathcal G} \gamma_x {\mathcal G},  
\end{equation} 
can be also used. As a result, we obtain Eq.~(\ref{Ogata2019}).
We used a similar relation in our previous paper \cite{Konye2020}.

\section{Explicit forms of $C^{(\ell)}_{abc}$ and $D^{(\ell)}_{abcd}$}

The summation over the Matsubara frequency in Eq.~(\ref{eq:DefDabcd})
can be carried out using complex integration (for details see the supplemental material of Ref.~\cite{Konye2020}).
For example,
\begin{equation}\begin{split}
&-k_{\rm B}T \sum_{n} 
\left(i\varepsilon_n + \frac{i\omega_\lambda}{2} \right)^\ell
({\mathcal G}_+)_a ({\mathcal G})_b ({\mathcal G})_c \cr
&=\int \frac{dz}{2\pi i} F(z) \biggl( z+\frac{i\omega_\lambda}{2} \biggr)^\ell
\mathcal G_a(z+i\omega_\lambda) \mathcal G_b(z) \mathcal G_c(z) \cr
&=\int \frac{d\varepsilon}{2\pi i} f(\varepsilon) \bigg[
\biggl( \varepsilon - \mu +\frac{i\omega_\lambda}{2} \biggr)^\ell
G_a^{\rm R}(\varepsilon+i\omega_\lambda) \cr 
&\times \{ 
G_b^{\rm R}(\varepsilon)G_c^{\rm R}(\varepsilon)
-G_b^{\rm A}(\varepsilon)G_c^{\rm A}(\varepsilon)\} 
+\biggl( \varepsilon -\mu -\frac{i\omega_\lambda}{2} \biggr)^\ell \cr
&\times \{ G_a^{\rm R}(\varepsilon)-G_a^{\rm A}(\varepsilon) \} 
G_b^{\rm A}(\varepsilon-i\omega_\lambda)G_c^{\rm A}(\varepsilon-i\omega_\lambda) \biggr],
\end{split}\end{equation}
where $\bm k$ in the thermal Green's functions is not shown and 
$F(z)=1/(e^{\beta z}+1)$ and $f(\varepsilon)$ is the Fermi distribution function 
with chemical potential $\mu$, 
$f(\varepsilon) = 1/(e^{\beta(\varepsilon-\mu)}+1)$.
The retarded (advanced) Green's function $G_a^{{\rm R}}$ ($G_a^{{\rm A}}$) 
is defined by
\begin{equation}
G_a^{{\rm R}} = \frac{1}{\varepsilon-\varepsilon_a ({\bm k})+i\Gamma},
\qquad G_a^{{\rm A}}= \left[G_a^{{\rm R}}\right]^*.
\end{equation}
After the analytic continuation of $i\omega_\lambda \rightarrow \hbar(\omega+i\delta$),
we can take the linear order with respect to $\omega$, which yields
\begin{equation}\begin{split}
C^{(\ell)}_{abc} =&\int \frac{d\varepsilon}{2\pi i} f(\varepsilon) \cr & \bigg[
( \varepsilon - \mu)^\ell
\left\{ \frac{\partial G_a^{\rm R}(\varepsilon)}{\partial \varepsilon} 
\left[ G_b^{\rm R}(\varepsilon)G_c^{\rm R}(\varepsilon)
-G_b^{\rm A}(\varepsilon)G_c^{\rm A}(\varepsilon)\right] \right.\cr
&\left.-\left[ G_a^{\rm R}(\varepsilon)-G_a^{\rm A}(\varepsilon) \right] 
\frac{\partial G_b^{\rm A}(\varepsilon)G_c^{\rm A}(\varepsilon)}
{\partial \varepsilon} \right\}\cr
&+\frac{\ell}{2}(\varepsilon-\mu)^{\ell-1}
\left\{ G_a^{\rm R}(\varepsilon)
\left[ G_b^{\rm R}(\varepsilon)G_c^{\rm R}(\varepsilon)
-G_b^{\rm A}(\varepsilon)G_c^{\rm A}(\varepsilon)\right] \right.\cr
&-\left[ G_a^{\rm R}(\varepsilon)-G_a^{\rm A}(\varepsilon) \right] 
G_b^{\rm A}(\varepsilon)G_c^{\rm A}(\varepsilon) \bigr\} \bigg].
\label{eq:Cabc}
\end{split}\end{equation}

Using the same method, we obtain
$\tilde C^{(\ell)}_{abc}=-\left[C^{(\ell)}_{cba}\right]^*$. 

The quantity $D^{(\ell)}_{abcd}$ can be calculated in a similar way, which gives
\begin{equation}\begin{split}
D^{(\ell)}_{abcd} =&\int \frac{d\varepsilon}{2\pi i} f(\varepsilon) \cr & \bigg[
( \varepsilon - \mu)^\ell
\biggl\{ \frac{\partial G_a^{\rm R}(\varepsilon)}{\partial \varepsilon} 
\bigl[ G_b^{\rm R}(\varepsilon)G_c^{\rm R}(\varepsilon)G_d^{\rm R}(\varepsilon) \cr
&-G_b^{\rm A}(\varepsilon)G_c^{\rm A}(\varepsilon)G_d^{\rm A}(\varepsilon) \bigr] \cr
&-\left[ G_a^{\rm R}(\varepsilon)-G_a^{\rm A}(\varepsilon) \right] 
\frac{\partial G_b^{\rm A}(\varepsilon)G_c^{\rm A}(\varepsilon)G_d^{\rm A}(\varepsilon)}
{\partial \varepsilon} \biggr\}\cr
&+\frac{\ell}{2}(\varepsilon-\mu)^{\ell-1}
\bigl\{ G_a^{\rm R}(\varepsilon)
\bigl[ G_b^{\rm R}(\varepsilon)G_c^{\rm R}(\varepsilon)G_d^{\rm R}(\varepsilon) \cr
&-G_b^{\rm A}(\varepsilon)G_c^{\rm A}(\varepsilon)G_d^{\rm A}(\varepsilon)\bigr] \cr
&-\left[ G_a^{\rm R}(\varepsilon)-G_a^{\rm A}(\varepsilon) \right] 
G_b^{\rm A}(\varepsilon)G_c^{\rm A}(\varepsilon)G_d^{\rm A}(\varepsilon) \bigr\} \bigg].
\label{eq:Dabc}
\end{split}\end{equation}
Similarly, we obtain
$\tilde D^{(\ell)}_{abcd}=-\left[D^{(\ell)}_{dcba}\right]^*$. 

\section{Small $\Gamma$ expansion of the energy integrals and detailed expressions for the electric and thermoelectric conductivity tensors}

We evaluate the energy integral in Eq.~(\ref{eq:Cabc})
in powers of $\Gamma^\ell$ in the small-$\Gamma$ region. 
We always have integrals of the form 
\begin{equation}
\int \frac{d\varepsilon}{2\pi i} f(\varepsilon) (\varepsilon-\mu)^\ell 
F(G_{a'}^{\rm R}, G_{a'}^{\rm A}, \cdots) 
(G_a^{\rm R}\cdots G_b^{\rm R}-G_a^{\rm A} \cdots G_b^{\rm A}).
\label{eq:GenInt}
\end{equation}
We calculate this kind of integrals using the contour integrals of complex functions. 
The case with $\ell=0$ was discussed before in Ref.~\cite{Konye2020}, and here, we extend the previous results to $\ell=1,2$.

First, there are poles at $\varepsilon= \mu + i(2n+1) k_{\rm B}T$ in $f(\varepsilon)$ 
with $n$ being an integer. 
However, the residues of these poles vanish as $\Gamma\rightarrow 0$ since 
the factor 
$G_a^{\rm R}\cdots G_b^{\rm R}-G_a^{\rm A} \cdots G_b^{\rm A}$
in (\ref{eq:GenInt}) approaches 0 as $\Gamma\rightarrow 0 $. 
This means that these residues are in the order of $\Gamma^1$.
Second, the factor $(\varepsilon-\mu)^\ell$ does not have singular points.
Third, the contribution containing only the retarded Green's functions
does not give any terms with $\Gamma^{-1}$ or $\Gamma^{-2}$, because even if the 
two Green's functions have the same energy $\varepsilon_a$, they only give 
higher order derivatives of $f(\varepsilon)$. 
Therefore, the contributions that give $\Gamma^{-1}$ or $\Gamma^{-2}$ are 
those that contain at least one $G_a^{\rm R}$ and one $G_a^{\rm A}$, which have the 
same energy $\varepsilon_a$ but have different imaginary parts, $\pm i\Gamma$.

Taking account of this prescription, Eq.~(\ref{eq:Cabc}) can be evaluated as
\begin{equation}
C^{(\ell)}_{abc} =\int \frac{d\varepsilon}{2\pi i} f'(\varepsilon) 
(\varepsilon - \mu)^\ell
G_a^{\rm R}(\varepsilon) G_b^{\rm A}(\varepsilon)G_c^{\rm A}(\varepsilon) +O(\Gamma^0),
\end{equation}
where we have used the integration by parts.
Then, if we make the contour integral in the lower half of the complex plane, 
only the pole of $G_a^{\rm R}$ contributes to the integral which gives
\begin{equation}
C^{(\ell)}_{abc} =-\frac{f'(\varepsilon_a-i\Gamma) (\varepsilon_a - \mu -i\Gamma)^\ell}
{(\varepsilon_a-\varepsilon_b -2i\Gamma)(\varepsilon_a-\varepsilon_c -2i\Gamma)} 
+O(\Gamma^0).
\label{eq:Cabc2}
\end{equation}
For example, we obtain
\begin{equation}\begin{split}
C^{(0)}_{aaa} &= -\frac{f'(\varepsilon_a-i\Gamma)}{(-2i\Gamma)^2} +O(\Gamma^0)\cr
&=\frac{f'(\varepsilon_a)}{4\Gamma^2} -i\frac{f''(\varepsilon_a)}{4\Gamma}+O(\Gamma^0).
\label{eq:Caaa}
\end{split}\end{equation}
Note that, in this case, we have used the $\Gamma$-expansion of the function 
$f(\varepsilon_a-i\Gamma)$. 
This treatment of the Fermi distribution function is unusual, but here we regard 
$f(\varepsilon)$ just as a mathematical complex function. 
Furthermore, note that, if we make the contour integral in the upper half 
of the complex plane for
\begin{equation}
C^{(0)}_{aaa} =\int \frac{d\varepsilon}{2\pi i} f'(\varepsilon) 
G_a^{\rm R}(\varepsilon) [G_a^{\rm A}(\varepsilon)]^2 +O(\Gamma^0),
\end{equation}
and take the residue of $[G_a^{\rm A}(\varepsilon)]^2$, we obtain the same result 
as in Eq.~(\ref{eq:Caaa}).

In a similar way, we obtain
\begin{equation}\begin{split}
C^{(1)}_{aaa} &=(\varepsilon_a-\mu) \left\{
\frac{f'(\varepsilon_a)}{4\Gamma^2} -i\frac{f''(\varepsilon_a)}{4\Gamma}\right\}
-i\frac{f'(\varepsilon_a)}{4\Gamma} +O(\Gamma^0), \cr
C^{(2)}_{aaa} &=(\varepsilon_a-\mu)^2 \left\{
\frac{f'(\varepsilon_a)}{4\Gamma^2} -i\frac{f''(\varepsilon_a)}{4\Gamma}\right\}
-i(\varepsilon_a-\mu)\frac{f'(\varepsilon_a)}{2\Gamma} \cr
&+O(\Gamma^0).
\end{split}\end{equation}
We can rewrite the above results as
\begin{equation}
C^{(\ell)}_{aaa} =
\frac{(\varepsilon_a-\mu)^\ell f'(\varepsilon_a)}{4\Gamma^2} 
-i\frac{\frac{\partial}{\partial\varepsilon_a} 
[(\varepsilon_a-\mu)^\ell f'(\varepsilon_a)]}{4\Gamma}+O(\Gamma^0),
\label{eq:GexpforC1}
\end{equation}
for $\ell=0,1,2$. 
Similarly, $C^{(\ell)}_{aab}$ and $C^{(\ell)}_{aba}$ can be calculated as
\begin{equation}
C^{(\ell)}_{aab} =C^{(\ell)}_{aba} = 
-i\frac{(\varepsilon_a-\mu)^\ell f'(\varepsilon_a)}
{2\Gamma (\varepsilon_a-\varepsilon_b)} +O(\Gamma^0).
\label{eq:GexpforC2}
\end{equation}
From Eq.~(\ref{eq:Cabc2}), we can see that all the others, such as 
$C^{(\ell)}_{abb}$ and $C^{(\ell)}_{abc}$ are at least 
in the order of $O(\Gamma^0)$. 

In a very similar way, $D^{(\ell)}_{abcd}$ can be evaluated from 
\begin{equation}
D^{(\ell)}_{abcd} = -\frac{f'(\varepsilon_a-i\Gamma)(\varepsilon_a - \mu -i\Gamma)^\ell}
{(\varepsilon_a - \varepsilon_b -2i\Gamma)(\varepsilon_a - \varepsilon_c -2i\Gamma)
(\varepsilon_a - \varepsilon_d -2i\Gamma)}+O(\Gamma^0).
\end{equation}
\begin{widetext}

From this expression, we obtain
\begin{equation}\begin{split}
D^{(\ell)}_{aaaa} =& (\varepsilon_a-\mu)^\ell 
\left\{ i\frac{f'(\varepsilon_a)}{8\Gamma^3} 
+ \frac{f''(\varepsilon_a)}{8\Gamma^2} - i\frac{f'''(\varepsilon_a)}{16\Gamma} \right\}
+\ell (\varepsilon_a-\mu)^{\ell-1} \left\{ \frac{f'(\varepsilon_a)}{8\Gamma^2} 
-i \frac{f''(\varepsilon_a)}{8\Gamma} \right\}
-i\frac{f'(\varepsilon_a)}{8\Gamma}\delta_{\ell,2} +O(\Gamma^0) \cr
=& i\frac{(\varepsilon_a-\mu)^\ell f'(\varepsilon_a)}{8\Gamma^3} 
+ \frac{\frac{\partial}{\partial\varepsilon_a}
[(\varepsilon_a-\mu)^\ell f'(\varepsilon_a)]}{8\Gamma^2} 
- i\frac{\frac{\partial^2}{\partial\varepsilon_a^2} 
[(\varepsilon_a-\mu)^\ell f'(\varepsilon_a)]}{16\Gamma} +O(\Gamma^0), \cr
D^{(\ell)}_{aaab} =& D^{(\ell)}_{aaba}=D^{(\ell)}_{abaa} 
= \frac{(\varepsilon_a-\mu)^\ell f'(\varepsilon_a)}{4\Gamma^2 (\varepsilon_a-\varepsilon_b)}
-i\frac{\frac{\partial}{\partial\varepsilon_a}
[(\varepsilon_a-\mu)^\ell f'(\varepsilon_a)]}{4\Gamma (\varepsilon_a-\varepsilon_b)}
+i\frac{(\varepsilon_a-\mu)^\ell f'(\varepsilon_a)}{2\Gamma (\varepsilon_a-\varepsilon_b)^2} +O(\Gamma^0), \cr
D^{(\ell)}_{aabb} =& D^{(\ell)}_{abba}=D^{(\ell)}_{abab} 
= -i \frac{(\varepsilon_a-\mu)^\ell f'(\varepsilon_a)}
{2\Gamma (\varepsilon_a-\varepsilon_b)^2} +O(\Gamma^0), \cr
D^{(\ell)}_{aabc} =& D^{(\ell)}_{abac}=D^{(\ell)}_{abca} 
= -i\frac{(\varepsilon_a-\mu)^\ell f'(\varepsilon_a)}
{2\Gamma(\varepsilon_a-\varepsilon_b)(\varepsilon_a-\varepsilon_c)} +O(\Gamma^0), 
\label{eq:GexpforD}
\end{split}\end{equation}
and all the other $D^{(\ell)}_{abcd}$ are in the order of $O(\Gamma^0)$.

Using the above expansions and 
substituting the matrix elements in Eq.~(\ref{eq:gammaDef}) 
for $\gamma_\mu$, $\gamma_\nu$.\, 
the electric and thermoelectric conductivity tensors are given as in 
Eq.~(\ref{eq:LmunuTotal}). 
Each contribution in Eq.~(\ref{eq:LmunuTotal}) is as follows:

\begin{equation}
L^{(\ell)C,1}_{\mu\nu} = \frac{e^{3-\ell} B}{mV} \sum_{{\bm k},a} 
\frac{(\varepsilon_a-\mu)^\ell f'(\varepsilon_a)}{4\Gamma^2} 
\left( \delta_{\mu y} \Eelln \Eellx + \delta_{\nu x} \Eellm \Eelly \right),
\label{eq:LC1munu}
\end{equation}
\begin{equation}
L^{(\ell)C,2}_{\mu\nu} = \frac{e^{3-\ell} B}{mV} \sum_{{\bm k},a,b (a\ne b)} {\rm Im} \biggl[
\frac{(\varepsilon_a-\mu)^\ell f'(\varepsilon_a)}{2\Gamma (\varepsilon_a- \varepsilon_b)} 
\left( \delta_{\mu y} p_{ab, \nu} p_{ba, x} + \delta_{\nu x} p_{ba, \mu} p_{ab, y} \right) \biggr],
\end{equation}
\begin{equation}
L^{(\ell),1}_{\mu\nu} = \frac{e^{3-\ell} B}{\hbar^2 V} \sum_{{\bm k},a} 
\frac{\frac{\partial}{\partial\varepsilon} 
[(\varepsilon_a-\mu)^\ell f'(\varepsilon_a)]}{4\Gamma^2} 
\Eellm \Eelln \Eellx \Eelly,
\label{eq:L1munu}
\end{equation}
\begin{equation}\begin{split}
L^{(\ell),2}_{\mu\nu} = \frac{e^{3-\ell} B}{\hbar^2 V} \sum_{{\bm k},a,b (a\ne b)} {\rm Re}
\biggl[& \biggl\{  
\frac{(\varepsilon_a-\mu)^\ell f'(\varepsilon_a)}{2\Gamma^2 (\varepsilon_a - \varepsilon_b)}
-i\frac{\frac{\partial}{\partial\varepsilon_a} 
[(\varepsilon_a-\mu)^\ell f'(\varepsilon_a)]}{2\Gamma (\varepsilon_a - \varepsilon_b)} 
+i\frac{(\varepsilon_a-\mu)^\ell f'(\varepsilon_a)}{\Gamma (\varepsilon_a - \varepsilon_b)^2}
\biggr\} \cr
&\times \bigl( \Eellm p_{ab,\nu} p_{ba,x} \Eelly 
+ \Eellm \Eelln p_{ab,x} p_{ba,y} + p_{ba,\mu} \Eelln \Eellx p_{ab,y}   \bigr)\biggr],
\label{eq:L2munu}
\end{split}\end{equation}
\begin{equation}
L^{(\ell),3}_{\mu\nu} = \frac{e^{3-\ell} B}{\hbar^2 V} \sum_{{\bm k},a,b (a\ne b)} {\rm Im}\biggl[
\frac{(\varepsilon_a-\mu)^\ell f'(\varepsilon_a)}{\Gamma (\varepsilon_a - \varepsilon_b)^2} 
( \Eellm p_{ab,\nu} \partial_x \varepsilon_b p_{ba,y}
+p_{ba,\mu} \Eelln p_{ab,x} \partial_y \varepsilon_b ) \biggr],
\end{equation}
\begin{equation}
L^{(\ell),4}_{\mu\nu} = \frac{e^{3-\ell} B}{\hbar^2 V} \sum_{{\bm k},a,b,c}^{'} {\rm Im}\biggl[  
\frac{(\varepsilon_a-\mu)^\ell f'(\varepsilon_a)}{\Gamma (\varepsilon_a - \varepsilon_b)(\varepsilon_a-\varepsilon_c)} 
( \Eellm p_{ab,\nu} p_{bc, x} p_{ca, y} + p_{ca, \mu} \Eelln p_{ab,x} p_{bc,y} ) \biggr],
\end{equation}
\begin{equation}
L^{(\ell),5}_{\mu\nu} = \frac{e^{3-\ell} B}{\hbar^2 V} \sum_{{\bm k},a,b (a\ne b)} {\rm Im}\biggl[  
\frac{(\varepsilon_a-\mu)^\ell f'(\varepsilon_a)}{\Gamma (\varepsilon_a - \varepsilon_b)^2}
p_{ba,\mu} p_{ab,\nu} p_{ba, x}  p_{ab, y}\biggr],
\end{equation}
\begin{equation}
L^{(\ell),6}_{\mu\nu} = \frac{e^{3-\ell} B}{\hbar^2 V} \sum_{{\bm k},a,b,c}^{'} {\rm Im}\biggl[  
\frac{(\varepsilon_a-\mu)^\ell f'(\varepsilon_a)}{\Gamma (\varepsilon_a - \varepsilon_b)(\varepsilon_a-\varepsilon_c)}  
p_{ca, \mu} p_{ab,\nu} p_{ba,x}  p_{ac,y} \biggr],
\label{eq:L6munu}
\end{equation}
where the summation with prime $\sum^{'}$ means that all the band indices 
($a,b,c$ or $a,b,c,d$) are different with each other.
Schematic representation of the contributions to 
$L^{(\ell),1}_{\mu\nu}$-$L^{(\ell),6}_{\mu\nu}$ are shown in Fig.~\ref{Fig:01}. 
\end{widetext}

\section{The derivation of the generalized $f$-sum rules}

In this appendix, we derive the generalized $f$-sum rules in Eqs.~(\ref{fSumRule}) and (\ref{fSumRule2}). 
Using the definition of $p_{ab,\mu}$ in Eq.~(\ref{eq:gammaDef}), we have
\begin{equation}
\sum_{b, (b \ne a)} \frac{p_{ab, \mu} p_{ba, \nu}}{\Eell-\varepsilon_b} 
= -\sum_{b, (b \ne a)} (\Eell-\varepsilon_b) \langle a| \partial_\mu b \rangle
\langle b| \partial_\nu a \rangle.
\end{equation}
Since the term with $b=a$ is equal to zero on the right-hand-side, we can add 
the $b=a$ term, resulting in the complete summation over $b$. 
Then, using the relation 
$\langle a| \partial_\mu b \rangle= -\langle \partial_\mu a|b \rangle$, 
since $\langle a | b\rangle = \delta_{ab}$, we can obtain
\begin{equation}\begin{split}
\sum_{b, (b \ne a)} \frac{p_{ab, \mu} p_{ba, \nu}}{\Eell-\varepsilon_b} 
&= \sum_{b}  \langle \partial_\mu a| b \rangle (\Eell-\varepsilon_b)
\langle b| \partial_\nu a \rangle \cr
&=\langle \partial_\mu a|(\Eell-H_{\bm k})| \partial_\nu a \rangle.
\end{split}\end{equation}
Now, it is apparent that the imaginary part of the right-hand-side is $iM_{\mu\nu}^a$.
The real part of the right-hand-side is obtained as follows.
Making the $k_\mu$ and $k_\nu$ derivative of $(\Eell-H_{\bm k})|a \rangle=0$,
we have
\begin{equation}\begin{split}
&\left( \partial_\mu \partial_\nu \Eell -\frac{\hbar^2}{m} \delta_{\mu\nu} \right) |a \rangle
+(\partial_\mu \Eell-\partial_\mu H_{\bm k} ) |\partial_\nu a \rangle \cr
&+(\partial_\nu \Eell-\partial_\nu H_{\bm k} ) |\partial_\mu a \rangle
+(\Eell-H_{\bm k} ) |\partial_\mu \partial_\nu a \rangle = 0.
\label{eq:SecondDeriv}
\end{split}\end{equation}
When we take the inner product using $\langle a |$, we obtain
\begin{equation}\begin{split}
&\partial_\mu \partial_\nu \Eell -\frac{\hbar^2}{m} \delta_{\mu\nu}\cr
&=-\langle a |(\partial_\mu \Eell-\partial_\mu H_{\bm k} ) |\partial_\nu a \rangle 
- \langle a| (\partial_\nu \Eell-\partial_\nu H_{\bm k} ) |\partial_\mu a \rangle \cr
&=\langle \partial_\mu a |(\Eell-H_{\bm k} ) |\partial_\nu a \rangle 
+\langle \partial_\nu a| (\Eell-H_{\bm k} ) |\partial_\mu a \rangle \cr
&=2 {\rm Re}\langle \partial_\mu a |(\Eell-H_{\bm k} ) |\partial_\nu a \rangle,
\label{eq:RealofM}
\end{split}\end{equation}
where we have used the relation 
\begin{equation}
(\partial_\mu \Eell-\partial_\mu H_{\bm k} ) | a \rangle 
=-(\Eell-H_{\bm k} ) |\partial_\mu a \rangle. 
\end{equation}
Combining these, we obtain the generalized $f$-sum rule in Eq.~(\ref{fSumRule}).

Next, we show another $f$-sum rule in Eq.~(\ref{fSumRule2}). 
Similarly, using the definition of $p_{ab,\mu}$ in Eq.~(\ref{eq:gammaDef}), we obtain
\begin{equation}
\sum_{b, (b \ne a)} \frac{p_{ab, \mu} p_{ba, \nu}}{(\Eell-\varepsilon_b)^2} 
= -\sum_{b, (b \ne a)} \langle a| \partial_\mu b \rangle
\langle b| \partial_\nu a \rangle.
\end{equation}
Then, the term with $b=a$ in the summation is added to use the complete set of $b$ and it is subtracted as follows:
\begin{equation}\begin{split}
\sum_{b, (b \ne a)} \frac{p_{ab, \mu} p_{ba, \nu}}{(\Eell-\Eellb)^2} 
&= \langle \partial_\mu a | \partial_\nu a \rangle 
- \langle \partial_\mu a | a \rangle \langle a | \partial_\nu a \rangle \cr
&={\rm Re} \langle \partial_\mu a | \partial_\nu a \rangle 
- \langle \partial_\mu a | a \rangle \langle a | \partial_\nu a \rangle
-\frac{i}{2} \ \Omega_{\mu\nu}^a,
\end{split}\end{equation}
where we have divided $\langle \partial_\mu a | \partial_\nu a \rangle$ into its real part and imaginary part. 

\section{Detailed calculations for obtaining the final results in Eq.~(\ref{eq:LmunuFinal})}

The remaining contributions are calculated as follows. Using the generalized 
$f$-sum rule, we obtain
\begin{equation}
L^{(\ell)C,2}_{\mu\nu} = \frac{e^{3-\ell} B}{mV} \sum_{{\bm k},a} 
\frac{(\varepsilon_a-\mu)^\ell f'(\varepsilon_a)}{2\Gamma} 
\left( \delta_{\mu y} M_{\nu x}^a + \delta_{\nu x} M_{\mu y}^a \right),
\end{equation}
and
\begin{equation}\begin{split}
&L^{(\ell),5}_{\mu\nu} + L^{(\ell),6}_{\mu\nu}
= \frac{e^{3-\ell} B}{\hbar^2 V} \sum_{{\bm k},a}
\frac{(\varepsilon_a-\mu)^\ell f'(\varepsilon_a)}{2\Gamma} \cr
&\times\biggl[ 
\left( \partial_\mu \partial_y \Eell - \frac{\hbar^2}{m} \delta_{\mu y} \right) M_{\nu x}^a
+\left( \partial_\nu \partial_x \Eell - \frac{\hbar^2}{m} \delta_{\nu x} \right) M_{y \mu}^a
\biggr].
\end{split}\end{equation}
Here, the Kronecker's delta functions in $L^{(\ell),5}_{\mu\nu} + L^{(\ell),6}_{\mu\nu}$
exactly cancel with $L^{(\ell)C,2}_{\mu\nu}$. 
Noting that the summation over $c$ in $L^{(\ell)C,4}_{\mu\nu}$ does not contain 
the terms with $c=a$ and $c=b$, we obtain
\begin{widetext}

\begin{equation}\begin{split}
L^{(\ell),4}_{\mu\nu} 
&= -\frac{e^{3-\ell} B}{\hbar^2 V} \sum_{{\bm k},a,b,c}^{'} 
\frac{(\varepsilon_a-\mu)^\ell f'(\varepsilon_a)}{\Gamma} {\rm Im}\biggl[  
\Eellm (\varepsilon_c- \varepsilon_b)
\langle a |\partial_\nu b \rangle \langle b |\partial_x c \rangle
\langle c |\partial_y a \rangle
+ (2{\rm nd}\ {\rm term}) \biggr] \cr
&= \frac{e^{3-\ell} B}{\hbar^2 V} \sum_{{\bm k},a,b (b\ne a),c (c\ne a)}  
\frac{(\varepsilon_a-\mu)^\ell f'(\varepsilon_a)}{\Gamma} {\rm Im}\biggl[ 
\Eellm \langle a |\partial_\nu b \rangle \langle \partial_x b |c \rangle
(H_{\bm k}- \varepsilon_b)
\langle c |\partial_y a \rangle + (2{\rm nd}\ {\rm term}) \biggr] \cr
&= \frac{e^{3-\ell} B}{\hbar^2 V} \sum_{{\bm k},a,b (b\ne a)} 
\frac{(\varepsilon_a-\mu)^\ell f'(\varepsilon_a)}{\Gamma} {\rm Im}\biggl[  
\Eellm \langle a |\partial_\nu b \rangle \left\{
\langle \partial_x b |(H_{\bm k}- \varepsilon_b)|\partial_y a \rangle
-\langle \partial_x b |a \rangle(\Eell- \varepsilon_b)\langle a |\partial_y a \rangle
\right\} + (2{\rm nd}\ {\rm term}) \biggr] \cr
&= \frac{e^{3-\ell} B}{\hbar^2 V} \sum_{{\bm k},a,b (b\ne a)} 
\frac{(\varepsilon_a-\mu)^\ell f'(\varepsilon_a)}{\Gamma} {\rm Im}\biggl[  
\Eellm \langle a |\partial_\nu b \rangle \left\{
\langle b |(\partial_x \varepsilon_b - \partial_x H_{\bm k})|\partial_y a \rangle
-\langle \partial_x b |a \rangle(\Eell- \varepsilon_b)\langle a |\partial_y a \rangle
\right\} + (2{\rm nd}\ {\rm term}) \biggr],
\end{split}\end{equation}
where the (2nd term) means that the contribution in which the variables 
$(\mu, \nu, x, y)$ in the first term are replaced by $(\nu, x, y,\mu)$.
We can see that the term containing $\partial_x \varepsilon_b$ in the last line 
exactly cancels with that in $L^{(\ell),3}_{\mu\nu}$. 
Therefore, we can take the summation over $b$ in the total 
of $L^{(\ell),3}_{\mu\nu}$ and $L^{(\ell),4}_{\mu\nu}$ as follows.
\begin{equation}\begin{split}
L^{(\ell),3}_{\mu\nu} + L^{(\ell),4}_{\mu\nu}
&= \frac{e^{3-\ell} B}{\hbar^2 V} \sum_{{\bm k},a} 
\frac{(\varepsilon_a-\mu)^\ell f'(\varepsilon_a)}{\Gamma} {\rm Im}\biggl[  
\Eellm \left\{ 
\langle \partial_\nu a |\partial_x H_{\bm k}|\partial_y a \rangle
-\langle \partial_\nu a |a \rangle \langle a |\partial_x H_{\bm k}|\partial_y a \rangle
-\langle \partial_\nu a |(\Eell- H_{\bm k})|\partial_x a \rangle
\langle a |\partial_y a \rangle \right\} \cr 
&+ (2{\rm nd}\ {\rm term}) \biggr].
\end{split}\end{equation}
Considering that $\langle a |\partial_\mu a\rangle$ is pure imaginary and 
using the relation in Eq.~(\ref{eq:RealofM}), we can see
\begin{equation}\begin{split}
L^{(\ell),3}_{\mu\nu} + L^{(\ell),4}_{\mu\nu}
&= \frac{e^{3-\ell} B}{\hbar^2 V} \sum_{{\bm k},a} 
\frac{(\varepsilon_a-\mu)^\ell f'(\varepsilon_a)}{\Gamma} {\rm Im}\biggl[  
\Eellm \biggl\{ 
\langle \partial_\nu a |\partial_x H_{\bm k}|\partial_y a \rangle
-\frac{1}{2} \partial_x \partial_y \Eell \langle \partial_\nu a |a \rangle \cr
&-\frac{1}{2} \left( \partial_\nu \partial_x \Eell -\frac{\hbar^2}{m}\delta_{\nu x}
\right) \langle a |\partial_y a \rangle \biggr\} 
+\Eelln \biggl\{ 
\langle \partial_x a |\partial_y H_{\bm k}|\partial_\mu a \rangle
-\frac{1}{2} \left( \partial_\mu \partial_y \Eell -\frac{\hbar^2}{m} \delta_{\mu y}
\right) \langle \partial_x a |a \rangle 
-\frac{1}{2} \partial_x \partial_y \Eell \langle a |\partial_\mu a \rangle \biggr\} \biggr],
\label{eq:FinalL34}
\end{split}\end{equation}
where we have used the relation
\begin{equation}
{\rm Re} \langle a |\partial_\mu H_{\bm k}|\partial_\nu a \rangle
={\rm Re} \langle a |(\partial_\mu H_{\bm k}-\partial_\mu \Eell )|\partial_\nu a \rangle
={\rm Re} \langle \partial_\mu a |(\Eell- H_{\bm k})|\partial_\nu a \rangle
=\frac{1}{2} \left( \partial_\mu\partial_\nu \Eell -\frac{\hbar^2}{m} \delta_{\mu\nu} 
\right).
\end{equation}

The final form of Eq.~(\ref{eq:FinalL34}) can be rewritten in a compact form using
the following relation.
Taking the inner product of Eq.~(\ref{eq:SecondDeriv}) with $\langle \partial_\tau a|$, 
we obtain
\begin{equation}\begin{split}
0&=\left( \partial_\mu \partial_\nu \Eell -\frac{\hbar^2}{m} \delta_{\mu\nu} \right) 
\langle \partial_\tau a|a \rangle
+\langle \partial_\tau a|(\partial_\mu \Eell-\partial_\mu H_{\bm k} ) |\partial_\nu a \rangle 
+\langle \partial_\tau a|(\partial_\nu \Eell-\partial_\nu H_{\bm k} ) |\partial_\mu a \rangle
+\langle \partial_\tau a|(\Eell-H_{\bm k} ) |\partial_\mu \partial_\nu a \rangle \cr
&=\left( \partial_\mu \partial_\nu \Eell -\frac{\hbar^2}{m} \delta_{\mu\nu} \right) 
\langle \partial_\tau a|a \rangle
+\partial_\mu \left[ \langle \partial_\tau a|(\Eell-H_{\bm k} ) |\partial_\nu a \rangle \right]
- \langle \partial_\tau \partial_\mu a|(\Eell-H_{\bm k} ) |\partial_\nu a \rangle
+\langle \partial_\tau a|(\partial_\nu \Eell-\partial_\nu H_{\bm k} ) |\partial_\mu a \rangle.
\end{split}\end{equation}
By substracting the same equation with $\mu$ and $\tau$ exchanged and 
taking the imaginary part, we obtain
\begin{equation}\begin{split}
0&=\left( \partial_\mu \partial_\nu \Eell -\frac{\hbar^2}{m} \delta_{\mu\nu} \right) 
{\rm Im} \langle \partial_\tau a|a \rangle
-\left( \partial_\tau \partial_\nu \Eell -\frac{\hbar^2}{m} \delta_{\tau\nu} \right) 
{\rm Im} \langle \partial_\mu a|a \rangle
+\partial_\mu M_{\tau\nu}^a -\partial_\tau M_{\mu\nu}^a
-\partial_\nu \Eell \Omega_{\tau\mu}^a 
-2{\rm Im}\langle \partial_\tau a|\partial_\nu H_{\bm k}|\partial_\mu a \rangle.
\end{split}\end{equation}
When we apply this equation to Eq.~(\ref{eq:FinalL34}), we obtain
\begin{equation}
L^{(\ell),3}_{\mu\nu} + L^{(\ell),4}_{\mu\nu}
= \frac{e^{3-\ell} B}{\hbar^2 V} \sum_{{\bm k},a} 
\frac{(\varepsilon_a-\mu)^\ell f'(\varepsilon_a)}{2\Gamma} \left\{  
\Eellm \left( \partial_y M_{\nu x}^a -\partial_\nu M_{yx}^a
-\partial_x \Eell \Omega_{\nu y}^a \right) 
+\Eelln (\partial_\mu M_{xy}^a -\partial_x M_{\mu y}^a
-\partial_y \Eell \Omega_{x \mu}^a ) \right\}.
\end{equation}
When we sum all the terms and carry out the integration by parts, 
we obtain Eq.~(\ref{eq:LmunuFinal}).
\end{widetext}

\bibliography{references}

\end{document}